\documentclass[11pt]{article}
\usepackage{jheppub}
\usepackage{amsmath,amssymb,amsfonts,graphicx,slashed,color}
\usepackage{subcaption}
\usepackage{tikz}
\usepackage{tikz-3dplot}
\usepackage[utf8]{inputenc}
\usetikzlibrary{decorations.markings}
\usetikzlibrary{decorations.pathmorphing}

\newcommand{\be}{\begin{equation}}
\newcommand{\ee}{\end{equation}}
\newcommand{\bea}{\begin{eqnarray}}
\newcommand{\eea}{\end{eqnarray}}
\newcommand{\beq}{\begin{equation}}
\newcommand{\eeq}{\end{equation}}
\newcommand{\beqn}{\begin{eqnarray}}
\newcommand{\eeqn}{\end{eqnarray}}

\newcommand{\cO}{\mathcal{O}}

\usepackage{upgreek}

\title{Holographic Entanglement and Poincar\'e blocks \\ in three-dimensional flat space}

\author[a]{Eliot Hijano}
\author[b,c]{\!, Charles Rabideau}

\affiliation[\,a]{Department of Physics and Astronomy, University of British Columbia,\\
6224 Agricultural Road, Vancouver, B.C.\ V6T 1Z1, Canada.}
\affiliation[\,b]{David Rittenhouse Laboratory, University of Pennsylvania,\\
209 S.33rd Street, Philadelphia PA, 19104, U.S.A.}
\affiliation[\,c]{Theoretische Natuurkunde, Vrije Universiteit Brussel (VUB), and \\ International Solvay Institutes, Pleinlaan 2, B-1050 Brussels, Belgium}

\emailAdd{ehijano@phas.ubc.ca}
\emailAdd{rabideau@sas.upenn.edu}

\abstract{
We propose a covariant prescription to compute holographic entanglement entropy and Poincar\'e blocks (Global BMS blocks) in the context of three-dimensional Einstein gravity in flat space. We first present a prescription based on worldline methods in the probe limit, inspired by recent analog calculations in AdS/CFT. Building on this construction, we propose a full extrapolate dictionary and use it to compute holographic correlators and blocks away from the probe limit. 
}


\begin{document}

\tikzset{->-/.style={decoration={
  markings,
  mark=at position #1 with {\arrow{>}}},postaction={decorate}}}

\def \L {10}
\def \H {1.5*\L}

\tikzset{
    mark position/.style args={#1(#2)}{
        postaction={
            decorate,
            decoration={
                markings,
                mark=at position #1 with \coordinate (#2);
            }
        }
    }
}

\tikzset{
  pics/carc/.style args={#1:#2:#3}{
    code={
      \draw[pic actions] (#1:#3) arc(#1:#2:#3);
    }
  }
}

\tikzset{point/.style={insert path={ node[scale=2.5*sqrt(\pgflinewidth)]{.} }}}

\tikzset{->-/.style={decoration={
  markings,
  mark=at position #1 with {\arrow{>}}},postaction={decorate}}}

  \tikzset{-dot-/.style={decoration={
  markings,
  mark=at position #1 with {\fill[red] circle [radius=3pt,red];}},postaction={decorate}}} 

 \tikzset{-dot2-/.style={decoration={
  markings,
  mark=at position #1 with {\fill[blue] circle [radius=3pt,blue];}},postaction={decorate}}} 

    \definecolor{darkgreen}{RGB}{0,180,0}

    \definecolor{purple2}{RGB}{222,0,255}

 \tikzset{-dot3-/.style={decoration={
  markings,
  mark=at position #1 with {\fill[purple2] circle [radius=3pt,purple2];}},postaction={decorate}}} 

\tikzset{snake it/.style={decorate, decoration=snake}}


\maketitle

\parskip=10pt

\section{Introduction / Motivation}
The holographic principle  \cite{tHooft:1999rgb,Susskind:1994vu} relates a theory of quantum gravity in a volume of space-time to a field theory in a lower dimensional boundary. The most fruitful concrete version of this principle is the AdS/CFT correspondence \cite{Maldacena:1997re}, which relates a theory of quantum gravity in asymptotically anti-de Sitter space-time (AdS) with a conformal field theory (CFT) at the boundary of AdS. Other versions of the same principle include the dS/CFT \cite{Strominger:2001pn},  Kerr/CFT \cite{Guica:2008mu}, Warped AdS/CFT \cite{Detournay:2012pc},  the Lifshitz/Non-relativistic CFT correspondence \cite{Balasubramanian:2008dm}, and the higher spin gauge theory / CFT correspondence  \cite{Klebanov:2002ja,Giombi:2009wh}. 

In this note we study the case of flat holography, where a quantum field theory is expected to arise as a dual to quantum gravity in asymptotically flat geometries. Research in this direction started in the four dimensional case \cite{Strominger:2013jfa}, and has recently gained attention due to possible implications concerning the infamous information paradox \cite{Hawking:2016msc}. The three-dimensional case \cite{Barnich:2010eb,Bagchi:2012cy} provides with a testing ground for ideas concerning classical and quantum gravity in flat space-times. 

However, unlike the case of asymptotically anti-de Sitter space-times, a full fledged holographic dictionary is missing in flat holography. Currently the only entry in the dictionary relates the asymptotic symmetry group of Minkowski space with the symmetry of the boundary field theory. Whether such a theory can be dynamically realized is currently unknown. This is however enough to try to understand the duality at the level of kinematics. In this note, we will study the holographic interpretation of entanglement entropy, two-point and three-point correlation functions, and global blocks. We will describe a prescription which correctly computes these objects in the probe limit\footnote{This is the limit where the dimensions of the operators in the correlation functions are in the regime $1 \ll m \ll 1/G$, so that the geodesic approximation can be used but back reaction can be neglected.}. Building on the intuition provided by this limit, we will propose a full extrapolate dictionary and successfully compute correlators and blocks involving light operators. We will then argue how the probe computations arise as a WKB approximation. 

Preliminarily, we will discuss the entanglement entropy of a boundary interval. The holographic picture will serve as motivation to discuss the other kinematic invariants of the theory. Entanglement entropy measures the correlation structure of a quantum system. Computations in the field theory are generally a daunting task. An example where the calculation can be performed analytically is the case of two dimensional conformal field theories, where the existence of a infinite dimensional symmetry simplifies the problem greatly.  Studying entanglement becomes more complicated as the dimension of the field theory increases, due to the lack of an infinite dimensional symmetry to capitalize on. Two dimensional Galilean conformal field theories (GCFT's) have been proposed as holographic duals to three-dimensional theories of gravity in asymptotically flat space-times \cite{Bagchi:2010eg}. The symmetry algebra is a central extension of the asymptotic symmetry of three dimensional Minkowski space-time at null infinity. It is known as the three dimensional Bondi-Metzner-Sachs algebra (BMS$_3$), and it is infinite dimensional. One can thus hope that the same simplifications arising in the conformal field theory case also arise in BMS field theories. The entanglement entropy calculations in the field theory literature are based in the replica trick \cite{Bagchi:2014iea} or ``flat" extensions of the Cardy formula \cite{Bagchi:2012xr,Barnich:2012xq,Jiang:2017ecm}.

In the context of AdS/CFT holography, Ryu and Takayanagi (RT) proposed that entanglement entropy was holographically computed by the area of a co-dimension two bulk minimal surface \cite{Ryu:2006bv}. A proposal for a covariant generalization was put forward by Hubeny, Rangamani and Takayanagi \cite{Hubeny:2007xt} (HRT), and was proven in \cite{Dong:2016hjy}. In order to establish a connection between Einstein gravity in three dimensional asymptotically Minkowski space and a BMS$_3$ field theory, it is necessary to understand whether a generalization of the RT/HRT proposals exist in flat space. The authors of \cite{Jiang:2017ecm}  have recently computed holographic entanglement entropy of a single interval region ${\cal A}$ in a BMS$_3$ field theory living at null infinity. The prescription, which they dubded the Rindler method,  emulates the Casini-Huerta-Myers \cite{Casini:2011kv} calculation of entanglement entropy of ball shaped regions in a conformal field theory. In the CFT case, the causal development of the boundary region is conformally mapped to hyperbolic space, where the theory enjoys a thermal periodicity in time. As a consequence, entanglement entropy is associated to the thermal entropy of the transformed  theory. In the flat case, the boundary region ${\cal A}$ is mapped to a different manifold with a thermal cycle using a finite BMS transformation. The thermal entropy of the transformed theory can then be computed by assuming  a flat version of the Cardy formula. The calculation can also be extended into the bulk. The geometry dual to the transformed theory  is a flat space cosmology (FSC), which possesses  a co-dimension one horizon whose area computes entanglement entropy. The geometric picture in the untransformed theory is the length of a space-like geodesic connected to two null geodesics $\gamma_i$ falling from the boundary endpoints ${\partial_{i}{\cal A}}$. The main properties of the geodesics in this picture is that they are tangent to the bulk modular flow vector whose boundary value corresponds to the modular generator. In \cite{Jiang:2017ecm}, it was also noted that the space-like geodesic connecting the null lines $\gamma_i$ is extremal. A figure of this set-up can be found in Figure \ref{fig:WSEE} (a). It is worth noting that the first calculation of holographic entanglement in three dimensional flat space was put forward in \cite{Bagchi:2014iea}, where it was shown that holographic entanglement can be computed using Wilson lines, much like the AdS construction introduced in \cite{Ammon:2013hba}. Nevertheless, the construction in \cite{Jiang:2017ecm} is explicitly geometric, and similar to the RT/HRT proposals in AdS. 

In this note, we will extend on the covariant formulation of the same geometric picture. The prescription consists of finding an extremal path connecting the end-points of the boundary interval, in the spirit of \cite{Hubeny:2007xt}. In flat space, the only geodesics that reach the points $\partial_{i}{\cal A}$ on future null infinity are null lines. 

 Unlike the case in AdS/CFT, the geodesics falling from the different boundary end-points do not intersect in the bulk, which forces us to connect them with an additional geodesic line. Extremizing the total length of the resulting path yields a covariant answer, equivalent to the one found in \cite{Jiang:2017ecm}. In section \ref{sec:FSC} we will test this proposal in non-trivial asymptotically flat geometries. 

Our proposal mimics the HRT formula for holographic entanglement entropy in AdS$_3$. In this case, the entanglement of a boundary interval is computed by the length of a bulk extremal geodesic connecting the interval endpoints, displaced slightly with a spacelike regulator away from the boundary.  In the context of AdS$_3$, a prescription for computing holographic scalar global conformal blocks in the probe limit has recently been proposed \cite{Hijano:2015rla,Castro:2017hpx}.  It consists on extremizing the length of a network of geodesic segments attached to the boundary of Anti de-Sitter space. The obvious prescription in flat space is to extremize a network of geodesics connected to the operators at the boundary through null geodesics, much like one does to compute entanglement entropy. Each of the geodesics is weighted by the mass of the corresponding primary operator in the BMS field theory, as drawn in Figure \ref{fig:3pt}. The physical intuition this picture provides is that the extremal weighted length of the network is the on-shell action of massive particles propagating in flat space. Our proposal in terms of on-shell actions simply reads
\beq \label{eq:EGi}
\text{global BMS block}=\text{extr}\, \left(e^{-\sum_i \xi_i S^{(i)}_{\text{on-shell}}}  \right)\, ,
\eeq
where $\xi_i$ are the masses of the particles and $S^{(i)}$ are the corresponding on-shell actions. This prescription can also be used to compute the two-point and three-point functions of scalar operators in a BMS$_3$ field theory. In section \ref{sec:SPIN}, this proposal will be generalized to include propagation of particles with spin, following the work in \cite{Castro:2014tta}. The calculations in this work correctly reproduce the probe limit of low-point correlators and general Poincar\'e blocks (or global BMS$_3$ blocks) computed in the field theory in \cite{Bagchi:2017cpu}. 

Inspired by the construction of global blocks in the probe limit, in section \ref{sec:extrapolate} we will propose a flat version of the extrapolate dictionary, and argue how the geodesic network construction arises as a WKB approximation. We will argue that the holographic construction of boundary correlators consists of attaching a position space Feynman diagram to null lines falling from the boundary at the operator locations, and integrating the position of the legs over an affine parameter;
\begin{align}
\langle \cO_1(x_1) \cO_2(x_2) \ldots \rangle 
\sim \int_{\gamma_{x_1}} d\lambda_1 \int_{\gamma_{x_2}} d\lambda_2 \ldots 
\langle \Psi_1(\lambda_1) \Psi_2(\lambda_2) \ldots \rangle\, .
\end{align}
Here, $\lambda_i$ is  the affine parameter along the null geodesic $\gamma_{x_i}$ and $\Psi_i$ is the bulk field dual to the operator  $\cO_i$. The symbol $\sim$ is used because both sides can be rescaled by arbitrary constant factors as both the normalisation of the operators and the measure of an affine parametrisation can be rescaled.

We turn now to a detailed description of the theories on both sides of the holographic duality. 

\subsection{Flat gravity in 2+1 dimensions}\label{sec:GravityIntro}
We consider Einstein gravity on $2+1$ dimensional Minkowski spacetime. We work on Eddington-Fikelstein coordinates, for which the line element reads
\beq \label{eq:LineElement}
ds^2=-du^2-2 du dr +r^2 d\phi^2\, ,
\eeq
with retarded time $u=t-r$ and the cylindrical identification $\phi\sim\phi+2\pi$. The future boundary is at $r\rightarrow\infty $ with $u$ and $\phi$ fixed. The holographic coordinate is $r$, which renders the holographic direction null. The asymptotic symmetry group is spanned by the following killing vectors
\beq\label{eq:KillingVectors}
\begin{split}
\xi^u&=u\partial_{\phi}Y(\phi )+T(\phi ) \,, \\
\xi^{\phi}&=Y(\phi )-{{u}\over{r}}\partial^2_{\phi}Y(\phi)-{1\over r}\partial_{\phi }T(\phi ) \,, \\
\xi^r&=-{J\over{2r}}\partial_{\phi}\xi^u-r \partial_{\phi}\xi^{\phi} \,,
\end{split}
\eeq
with the restriction $(\partial^3_{\phi}+\partial_{\phi})T(\phi )=(\partial^3_{\phi}+\partial_{\phi})Y(\phi )=0$. Expanding $Y(\phi)$ and $T(\phi)$ in the modes $e^{i n\phi}$ with $n=-1,0,1$, the killing vectors realize a sub-algebra of the Galilean Conformal Algebra (GCA).
\beq\label{eq:ASG}
\begin{split}
[L_i,L_j]&=(i-j)L_{i+j} \,, \\
[L_i,M_j]&=(i-j)M_{i+j} \,, \\
[M_i,M_j]&=0 \,,
\end{split}
\eeq
where $i,j=-1,0,1$. At null infinity this algebra is enhanced to an infinitely dimensional asymptotic symmetry group known as the BMS$_3$ group, or also the 2d Galilean Conformal Algebra. The conserved charges of this group satisfy a centrally extended version of the algebra that reads
\beq\label{eq:CentralExtension}
\begin{split}
[{\cal L}_i,{\cal L}_j]&=(i-j){\cal L}_{i+j} +{{c_L}\over{12}}(i+1)i(i-1)\delta_{i+j,0}\,, \\
[{\cal L}_i,{\cal M}_j]&=(i-j){\cal M}_{i+j}+{{c_M}\over{12}}(i+1)i(i-1)\delta_{i+j,0} \,, \\
[{\cal M}_i,{\cal M}_j]&=0 \, .
\end{split}
\eeq
The central charges depend on the specific theory of gravity. For the case of Einstein gravity, in order for the phase space of
three-dimensional asymptotically flat gravity to match the space of coadjoint representations of the
BMS$_3$ group, it is required that \cite{Barnich:2015uva,Oblak:2016eij}
\beq
c_L=0\, , \quad c_M={3\over G}\, .
\eeq
Invoking the holographic principle in the semi-classical limit implies that a theory of gravity in asymptotically flat space-times is described by a field theory that enjoys BMS symmetry. The field content of such theory must organize into represenations of this symmetry; $\{L_0,M_0\}$ form a maximally commuting subalgebra, so we will label out representations by their eigenvalues $\{\Delta,\xi \}$. The generators $\{L_{-1},L_0,L_1,M_{-1},M_0,M_1\}$ form the global sub-algebra which corresponds to the Poincar\'e group. This sub-algebra has two quadratic casimirs \cite{Bagchi:2016geg}
\begin{align}
C_1 &= M_0^2 - M_{-1} M_1 \\
C_2 &= 2 L_0 M_0 - \frac12 \left( L_{-1} M_{1} +L_{1} M_{-1} +M_{1} L_{-1} +M_{-1} L_{1} \right)
 \,.
\end{align}
The eigenvalues of the casimirs are 
\begin{align}
\lambda_1 = \xi^2 \qquad \lambda_2 = 2 \xi (\Delta-1) \, .
\end{align}
In order to make the connection to bulk physics, it helps to write these in terms of the familiar generators of the Poincar\'e algebra  
\begin{align}
[P_\mu,P_\nu]&=0 
\qquad [\mathcal{M}_{\mu,\nu},P_\rho] 
= i \left(\eta_{\mu \rho}P_\nu -\eta_{\nu \rho} P_\mu  \right) \\
 [\mathcal{M}_{\mu,\nu},\mathcal{M}_{\rho,\sigma}]
   &= i \left( \eta_{\mu \rho} M_{\nu \sigma} - \eta_{\mu \sigma} M_{\nu \rho}
   -\eta_{\nu \rho} M_{\mu \sigma} + \eta_{\nu \sigma} M_{\mu \rho} \right) \,,
\end{align}
which are related by
 \begin{align}
 P_t &= M_0 &\quad P_x &= \frac{ M_1 + M_{-1} }{2} &\quad P_y &= i \frac{ M_1 - M_{-1} }{2} \,,\\
 \mathcal{M}_{xy} &=  L_0 &\quad  \mathcal{M}_{ty} &= \frac{ L_1 + L_{-1} }{2} &\quad  \mathcal{M}_{tx} &= -i \frac{ L_1 - L_{-1} }{2} \,.
 \end{align}
 In terms of these, the casimirs read
  \begin{align}
  C_1 = -P^2 
  \qquad C_2 = \epsilon^{\mu\nu\rho}  M_{\mu\nu} P_\rho \,.
  \end{align}
Matching these casimirs allows us to identify the mass as $m^2 = \xi^2$ and determine the spin in terms of $\Delta$. For a scalar field, $C_1$ imposes the Klein-Gordon equation whereas $M_{\mu\nu}$ acts trivially so that $C_2=0$ and $\Delta=1$.

  Note that in 3 dimensions, spin is a continuously tunable parameter. We conclude that BMS primary operators must correspond to massive spinning particles propagating in flat space. A conjecture for the extrapolate dictionary relating flat space quantum field solutions and local BMS primary operator insertions will be discussed in section \ref{sec:extrapolate}.

We now turn to a more detailed description of some properties of BMS$_3$ field theories.

\subsection{BMS$_3$ Field Theory}
The holographic principle applied to flat space relies on the existence of a field theory living at null infinity. The symmetry of the theory must consist of the  BMS$_3$  algebra. A theory which realises this symmetry is a BMS$_3$ Field Theory or BMSFT. It is currently not clear whether this algebra can be dynamically realized, as very few examples of such theories exist. For example, in \cite{Barnich:2012rz} the authors construct a prototypical theory with BMS$_3$ symmetry,  as a limit of Liouville theory. In \cite{Banerjee:2015kcx} other exampes are constructed using holomorphic free fields. We will simply study kinematics which are fixed by the symmetry properties of the theory. We consider an asymptotically flat bulk space-time ${\cal M}$. At null infinity the manifold is endowed with a degenerate metric
\beq\label{eq:MetricBoundary}
ds^2=0  du^2+d\phi^2\, ,
\eeq
where $\phi$ is the solid angle in $S^1$. The diffeomorphism group is the BMS$_3$ group, and it acts on the coordinates of the theory as
\beq\label{eq:FiniteBMS}
\begin{split}
u&\rightarrow u'=u \partial f(\phi)+g(\phi ),\\
\phi &\rightarrow \phi'=f(\phi)\, .
\end{split}
\eeq

As an aside, we should note that this transformation law means that any two points $(u_1,\phi_1)$ and $(u_2,\phi_2)$ with  $\phi_1 \neq \phi_2$ can be mapped to lie at the same $u=0$ by 
\begin{align}
f(\phi) = \phi   \,, \qquad  g(\phi) = \frac{u_2 \sin \left(\phi -\phi _1\right)-u_1 \sin \left(\phi -\phi
   _2\right)}{\sin\left(\phi _1-\phi _2\right)} \,.
\end{align}
By micro-causality, we must conclude that any local operators inserted at points such that $\phi_1 \neq \phi_2$ must commute. In other words, a BMS$_3$ invariant theory must be ultra-local with no propagating degrees of freedom. However, \cite{Jiang:2017ecm} found non-trivial entanglement in these theories. The simplest scenario consistent with these facts is a local theory of local degrees of freedom which do not propagate, but are subject to non-local constraints which impose non-trivial correlations.

 The correlators we are hoping to compute are kinematic invariants, and have been written on the plane in \cite{Bagchi:2017cpu}. The expressions for the two- and three-point functions in the cylinder can be obtained by performing the BMS transformation
\beq
x=e^{i\phi}\, ,\quad t=i u e^{i\phi}\, ,
\eeq
which acts on primaries as \cite{Bagchi:2013qva} $\Phi(u,\phi)=e^{i\phi \Delta}e^{i u \xi}\Phi(x,t)$. The result we would like to reproduce then reads
\beq\label{eq:lowpt}
\begin{split}
\langle \Phi_{\Delta,\xi} \Phi_{\Delta,\xi}\rangle& ={{e^{{-\xi{{u^{\partial}_2-u^{\partial}_1}\over{\tan{{\phi^{\partial}_2-\phi^{\partial}_1}\over 2}}}}}}} \left( \sin {{\phi^{\partial}_2-\phi^{\partial}_1}\over 2} \right)^{-2\Delta}\, , \\
\langle \Phi_{\Delta_1,\xi_1} \Phi_{\Delta_2,\xi_2}\Phi_{\Delta_3,\xi_3}\rangle&=C_{123}\prod_k{{e^{
\xi_k{{\sum_{i<j}(-1)^{1+i+j}(u^{\partial}_i-u^{\partial}_j)\cos(\phi^{\partial}_k-\phi^{\partial}_i)}
\over
{\sum_{i<j}(-1)^{1+i+j}\sin(\phi^{\partial}_i-\phi^{\partial}_j)}}
} }} \left( \sin {{\phi^{\partial}_i-\phi^{\partial}_j}\over 2} \right)^{-\Delta_{ijk}}\, ,
\end{split}
\eeq
where the primary operators are located in the null plane at points $(u,\phi)=(u^{\partial}_i,\phi^{\partial}_i)$, and have quantum numbers $\xi_i$ and $\Delta_i$. 

The four-point function is not fixed by BMS symmetry. It can however be expanded in a basis of BMS invariant functions, or global BMS$_3$ blocks. In this paper we study the block of four identical  operators $\Phi_{\Delta,\xi}$ that interchange a BMS$_3$ representation with primary $\Phi_{\Delta_p,\xi_p}$. We will place the four primaries at boundary locations $x_k=(u^{\partial}_k,\phi^{\partial}_k)$ for $k=1,2,3,4$, and the channel of the block will be $12\rightarrow 34$. The four-point function of BMS$_3$ primaries on the plane decomposes into blocks as
\beq\label{eq:4PtFunction}
\begin{split}
\langle \phi_1\phi_2\phi_3\phi_4\rangle &=\prod_{1\leq i \leq j\leq 4}x^{-\sum_k \Delta_{ijk}/3}_{ij}e^{{{t_{ij}}\over{x_{ij}}}\sum_k\xi_{ijk}/3}F_{BMS}(x,t) \, ,\\
F_{BMS}(x,t)&=(1-x)^{(\Delta_{231}+\Delta_{234})/3}x^{(\Delta_{341}+\Delta_{342})}e^{   {{t}\over{1-x}} {{\xi_{231}+\xi_{234}}\over 3} -3{{t}\over{x}}(\xi_{341}+\xi_{342})}  \, \\
&\times \sum_{p} C_{12p}C_{34p} g_{p}(x,t)\, ,
\end{split}
\eeq
where the crossratios are defined as
\beq\label{eq:CrossRatios}
x={{x_{12}x_{34}}\over{x_{13}x_{24}}} \, , \quad {t\over x}={{t_{12}}\over{x_{12}}}+{{t_{34}}\over{x_{34}}}-{{t_{13}}\over{x_{13}}}-{{t_{24}}\over{x_{24}}}\, .
\eeq
The constants $C_{12p}$ and $C_{34p}$ depend on the details of the field theory. 
In the large central charge limit, the full BMS block reduces to the global block. 
For equal external primary operators, the global block reads
\beq\label{eq:Block}
g_{p}(x,t)=2^{2\Delta_p-2}(1-x)^{-1/2}x^{\Delta_p-2\Delta}(1+\sqrt{1-x})^{2-2\Delta_p} e^{2\xi{t\over x}-\xi_p {{t}\over{x\sqrt{1-x}}}}\, .
\eeq

We turn now to the computation of holographic entanglement entropy, which will serve as motivation for the introduction of geodesic networks to compute low-point correlation functions and holographic global conformal blocks in the probe limit. 
 
\section{Flat space holographic Entanglement Entropy}
In this section we provide with an alternative covariant re-formulation of the geometrical picture presented in \cite{Jiang:2017ecm}. The background metric is given by equation \eqref{eq:LineElement}, and the boundary is taken to be null future infinity, at $r\rightarrow\infty$, with $u$ and $\phi$ fixed. We define an interval ${\cal A}$ as a straight line connecting two points at future infinity. We denote them by $\partial_i {\cal A}$ and they are located at $(u^{\partial}_i,\phi^{\partial}_i)$.  The authors of \cite{Jiang:2017ecm} compute holographic entanglement entropy as the length of a space-like geodesic which is connected to the interval at null infinity by two null geodesics. The space-like geodesic is the fixed points of replica symmetry, and the null geodesics are tangential to the modular flow. 
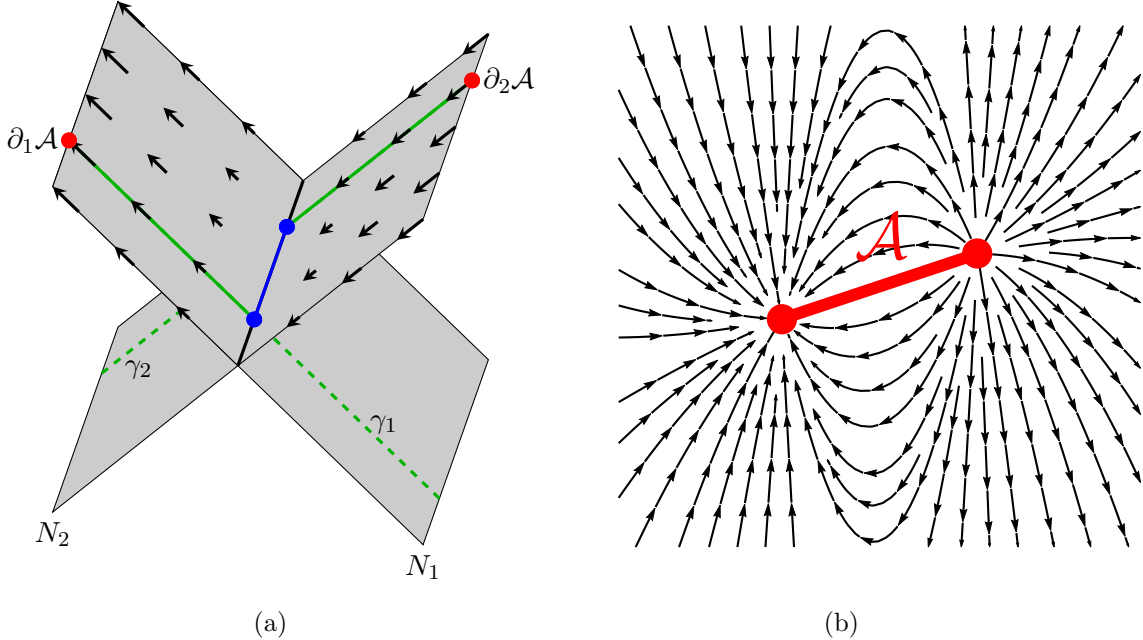
\begin{figure}[]

\centering
\begin{subfigure}[t]{0.48\textwidth}
        \centering
       \tdplotsetmaincoords{60}{100}
\begin{tikzpicture}[scale=5,tdplot_main_coords,vectorR/.style={-stealth,red,very thick},vectorB/.style={-stealth,blue,very thick},vectorP/.style={-stealth,black,very thick}]

    \def\x{1}

  \filldraw[
        draw=black,%
        fill=black!20,%
    ]          (0,1,0)
            -- (1,1,0)
            -- (1,1/2,1/2)
            -- (0,1/2,1/2)
            -- cycle;

  \draw[
        darkgreen,dashed,very thick%
    ]          (3/4,1,0)
            -- (3/4,1/2,1/2) node [pos=0.3,above,name=int] {{\textcolor{black}{$\gamma_1$}}};

    \filldraw[
        draw=black,%
        fill=black!20,%
    ]          (0,0,0)
            -- (1,0,0)
            -- (1,1,1)
            -- (0,1,1)
            -- cycle;

  \draw[
        darkgreen,dashed,very thick%
    ]          (1/4,0,0)
            -- (1/4,1/2,1/2) node [pos=0.2,below=2,name=int] {{\textcolor{black}{$\gamma_2$}}};

  \filldraw[
        draw=black,%
        fill=black!20,%
    ]          (0,0,1)
            -- (1,0,1)
            -- (1,1/2,1/2)
            -- (0,1/2,1/2)
            -- cycle;

  \filldraw[
        draw=black,%
        fill=black!20,%
    ]          (0,0,1)
            -- (1,0,1)
            -- (1,1/2,1/2)
            -- (0,1/2,1/2)
            -- cycle;

\draw (1,0,0) node[below] {$N_2$};
\draw (1,1,0) node[below] {$N_1$};

  \draw[
         darkgreen,dashed,very thick%
    ]          (3/4,0,1)
            -- (3/4,1/2,1/2) node [pos=0,left,name=O1] {\textcolor{black}{$\partial_1 {\cal A}$}};
  \draw[
          darkgreen,dashed,very thick%
    ]          (1/4,1,1)
            -- (1/4,1/2,1/2) node [pos=0,right,name=O2] {\textcolor{black}{$\partial_2 {\cal A}$}};

  \draw[
       black,very thick%
    ]          (0,1/2,1/2)
            -- (1,1/2,1/2);

  \draw[
         -dot-=0,darkgreen,very thick%
    ]          (3/4,0,1)
            -- (3/4,1/2,1/2);

  \draw[
        -dot-=0,darkgreen,very thick%
    ]          (1/4,1,1)
            -- (1/4,1/2,1/2);

  \draw[
       -dot2-=0, -dot2-=1,blue,very thick%
    ]       (3/4,1/2,1/2)
            -- (1/4,1/2,1/2);

\foreach \VN in {0,...,4}{
	\draw[vectorP]  (\VN/4,1,1) --   (\VN/4,{1-0.1/sqrt(2)},{1-0.1/sqrt(2)}) ;
}
\foreach \VN in {0,...,4}{
	\draw[vectorP]  (\VN/4,5/6,5/6) --   (\VN/4,{5/6-0.075/sqrt(2)},{5/6-0.075/sqrt(2)}) ;
}

\foreach \VN in {0,...,4}{
	\draw[vectorP]  (\VN/4,4/6,4/6) --   (\VN/4,{4/6-0.05/sqrt(2)},{4/6-0.05/sqrt(2)}) ;
}

\foreach \VN in {0,...,4}{
	\draw[vectorP]     (\VN/4,{0+0.1/sqrt(2)},{1-0.1/sqrt(2)}) --(\VN/4,0,1);
}
\foreach \VN in {0,...,4}{
	\draw[vectorP]     (\VN/4,{1/6+0.075/sqrt(2)},{5/6-0.075/sqrt(2)}) -- (\VN/4,1/6,5/6);
}

\foreach \VN in {0,...,4}{
	\draw[vectorP]    (\VN/4,{2/6+0.05/sqrt(2)},{4/6-0.05/sqrt(2)}) -- (\VN/4,2/6,4/6) ;
}

  \draw[
         -dot-=0,darkgreen,very thick%
    ]          (3/4,0,1)
            -- (3/4,0+0.01,1-0.01);

  \draw[
        -dot-=0,darkgreen,very thick%
    ]          (1/4,1,1)
            -- (1/4,1-0.01,1-0.01);

\end{tikzpicture}
\caption{ }
\end{subfigure}
\begin{subfigure}[t]{0.48\textwidth}
\centering
\begin{picture}(200,200)
\put(0,0){\includegraphics[scale=0.4]{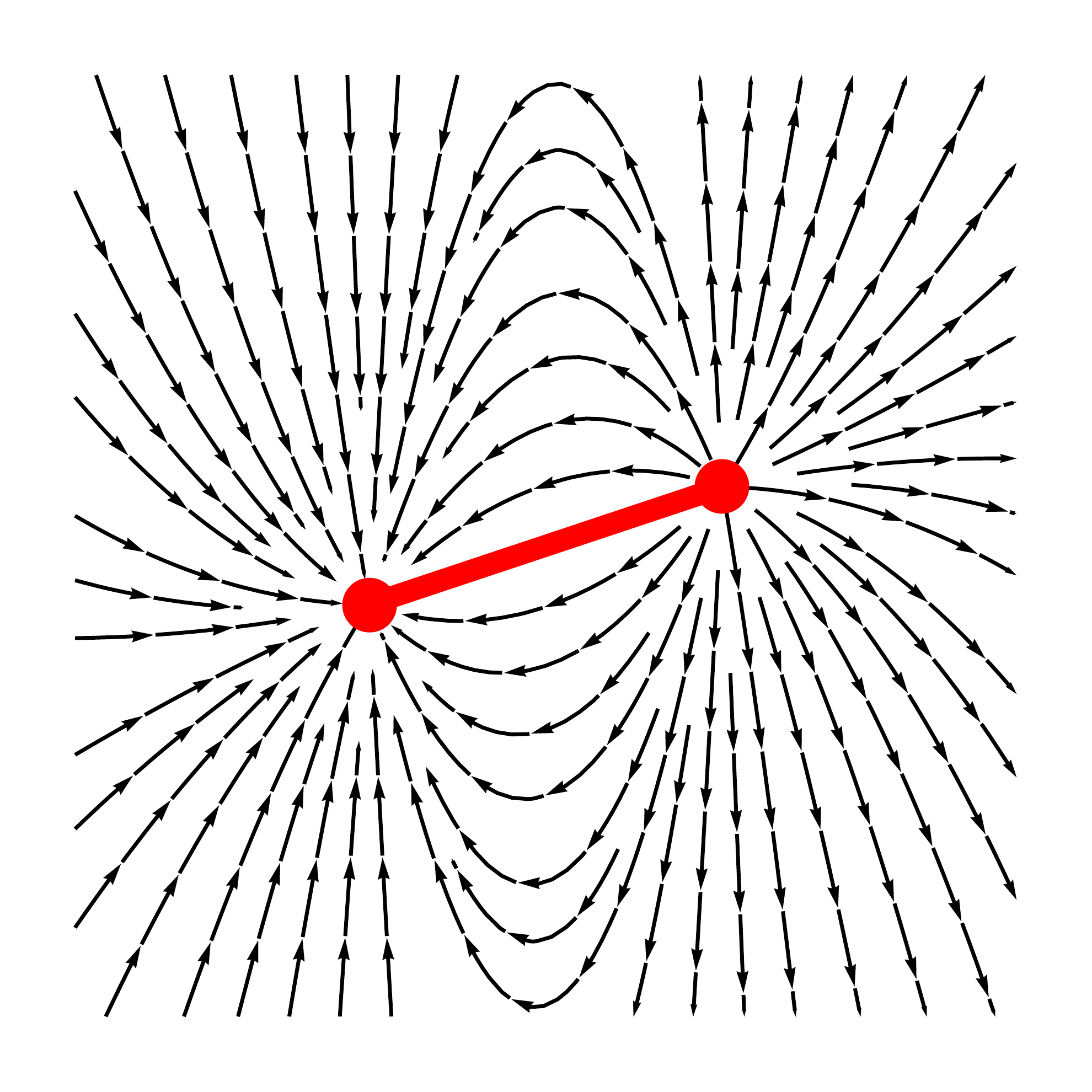}}
\put(105,125){{\Huge \textcolor{red}{${\cal A}$}}}
\end{picture}
\caption{ }
\end{subfigure}
    \caption{a) Holographic entanglement entropy as understood in \cite{Jiang:2017ecm}. The gray planes ($N_i$) correspond to the codimension one Cauchy Horizon of the flat space cosmological solution. The green lines are null, and we have dubbed them $\gamma_1$ and $\gamma_2$ throughtout this note. The black arrows stand for the modular flow vector in the bulk, which vanishes at the intersection between the two planes; $N_1\cap N_2$. The path is then formed by null lines tangent to modular flow (green) and the fixed points of replica symmetry (blue). b) Modular flow at the boundary. The axis are $\phi, u$. The red line is the boundary region ${\cal A}$ whose entanglement entropy is being studied. 
}\label{fig:WSEE}
\end{figure} 

This construction arises from an inverse bulk Rindler map, which is the BMS analog to the conformal map used in \cite{Casini:2011kv} to compute holographic entanglement entropy in the AdS/CFT context. The map transforms a flat space cosmological solution (FSC) back to the original Minkowski space-time. The Cauchy Horizon of the FSC maps to the union of two null planes in Minkowski space. The equations for these planes read
\beq\label{eq:originalNi}
N_i\, \text{:}\, \quad u=u^{\partial}_i-r+r\cos \left( \phi-\phi^{\partial}_i \right)\, .
\eeq
Each plane contains a null line $\gamma_i$ with $u=u^{\partial}_i$ and $\phi=\phi^{\partial}_i$. These lines are tangential to the bulk modular flow. Entanglement entropy is computed in \cite{Jiang:2017ecm} as the length of the space-like geodesic connecting $\gamma_1$ and $\gamma_2$ along $N_1\cap N_2$.  The construction is summarized in Figure \ref{fig:WSEE}.

We would like to re-phrase this prescription in the language of extremal geodesics connecting the boundary endpoints $\partial{\cal A}_i$, in the spirit of HRT \cite{Hubeny:2007xt}. For this we will first review the AdS case, which we will then generalize to the flat space case at hand. 

\subsection{Review of covariant holographic entanglement entropy in AdS/CFT}
Before we attempt to understand the flat space case, it is instructive to review the construction in AdS/CFT.  The vacuum entanglement entropy for a conformal field theory region in the plane consisting of an interval $[0,\omega]$ with $\omega\in\mathbb{C} $ is given by \cite{Calabrese:2004eu}
\beq\label{eq:AdSSEE}
S^{\text{CFT}_2}_{\text{EE}}={c\over 6}\log {{\omega}\over{\epsilon}}+{\bar{c}\over 6}\log {{\bar{\omega}}\over{\epsilon}}\, ,
\eeq
where $c$, $\bar{c}$ are the central charges of the conformal field theory and $\epsilon$ is the lattice distance, acting as a regulator of UV divergences.

\iftrue

For the holographic picture, we follow  \cite{Hubeny:2007xt} very closely, where this prescription was first presented. The RT prescription for calculating entanglement is usually associated to Euclidean geometries. In relies on finding minimal surfaces, which is not a well defined Lorentzian concept. Wiggling space-like surfaces in the time direction one can make the area of a surface arbitrarily small. The solution to this puzzle is to consider the extrema of the area of a surface instead of the minima.  We thus need to solve $\delta \text{Area}=0$ for small variations of the location of the co-dimension two surface, which we will call ${\cal S}$. We specify ${\cal S}$ with two constraints
\beq
\varphi_1(x^{\nu})=0\, ,\quad  \varphi_2(x^{\nu})=0\, .
\eeq
We construct two null vectors $N_{\pm}^{\mu}$ obeying $N_{+}^2=N_{-}^2=0$ and $N_{+}\cdot N_{-}=-1$, such that they are orthogonal to the surface. We thus write
\beq\label{eq:NullVectorsN}
N^{\mu}_{\pm}={\cal N}g^{\mu \nu}\left(\nabla_{\nu} \varphi_1+\mu_{\pm} \nabla_{\nu}\varphi_2    \right)\, ,
\eeq
and solve for $\mu_{\pm}$ and ${\cal N}$. The induced metric on ${\cal S}$ reads
\beq
h_{\mu\nu}=g_{\mu\nu}+N_{+\mu}N_{-\nu}+N_{+\nu}N_{-\mu}\, ,
\eeq
and the null curvatures can be written as
\beq
\left(\chi_{\pm}\right)_{\mu\nu}=h^{\rho}_{\, \, \mu}h_{\sigma \nu}\nabla_{\rho}N_{\pm}^{\sigma}\, .
\eeq
The null expansions are then defined as follows
\beq
\theta_{\pm}=g^{\mu\nu}\left(\chi_{\pm}\right)_{\mu\nu}\, .
\eeq
Finally, the variation of the area of the surface ${\cal S}$ under a variation of its location can be written in terms of the null expansions as
\beq
\delta \text{Area}\sim \int_{{\cal S}} \left( \theta_{+} N_{+}^{\mu}  \delta X_{\mu} + \theta_{-} N_{-}^{\mu}  \delta X_{\mu}    \right)\, .
\eeq
The variation of the area will vanish if $\theta_{\pm}=0$, rendering the surface ${\cal S}$ extremal. 

For the sake of clarity and simplicity, we will perform the computation of holographic entanglement entropy in pure AdS$_3$, where the answer must match the CFT$_2$ vacuum entanglement entropy of equation \eqref{eq:AdSSEE}. We consider the Poincar\'e metric
\beq
ds^2={{-dt^2+dx^2+dz^2}\over{z^2}}\, ,
\eeq
where the holographic boundary lives at $z\rightarrow 0$. Consider a general co-dimension two curve ${\cal S}$ specified by 
\beq
\varphi_1=t-T(z)\, \quad \varphi_2=x-X(z)\, .
\eeq
The computation of the null expansions yileds
\beq
\begin{split}
\theta_{+}-\theta_{-}&\sim \left( X'^3+X'(1-T'^2)-z X'' \right)\, , \\
\theta_{+}+\theta_{-}&\sim \left(   
T'^3+z(1+X'^2)T''-T'(1+X'^2+zX'X'')
\right)\, ,
\end{split}
\eeq
where the symbol $\sim$ omits non-vanishing dependence on $X(z)$ and $T(z)$. We now need to solve for $\theta_{\pm}=0$. The first equation implies
\beq\label{eq:Gp}
T'(z)=\pm \sqrt{{{X'+X'^3-z X''}\over{X'}}}\, ,
\eeq
which can be used in the second differential equation to obtain
\beq
X'(3X''+zX''')-3z X''^2=0\, .
\eeq
The solution reads
\beq
(X(z)-c_3)^2=c_1+c_2 z^2 \, ,
\eeq
where $c_i$ are arbitrary constants. This implies, in virtue of equation \eqref{eq:Gp},
\beq
(T(z)-d_1)^2=(c_1+c_2 z^2)\left(1+{1\over{c_2}}\right)\, ,
\eeq
where again $d_i$ are constants. We now need to fix the constants of integration by depanding that the curve ${\cal S}$ starts and ends at the asymptotic boundary at the points $(x_1,t_1)$ and $(x_2,t_2)$. After this, the solution for the curve reads
\beq 
\begin{split}
(X(z)-x_1)(X(z)-x_2)&={{z^2 (x_1-x_2)^2}\over{(t_1-t_2-x_1+x_2)^2}}\, , \\
(T(z)-t_1)(T(z)-t_2)&={{z^2 (t_1-t_2)^2}\over{(t_1-t_2-x_1+x_2)^2}}\, .
\end{split}
\eeq
The length of such curve from point $(x_1,t_1)$ and $(x_2,t_2)$ is formally infinite, but can be regulated by moving the holographic screen to $z=\epsilon\ll 1$. The regulated length is
\beq
S_{EE}={1\over {4G}}L^{\text{extremal}}_{{\cal S}} ={1\over{4G}} \left( \log{{\omega_{12}}\over{\epsilon}}+\log{{\bar{\omega}_{12}}\over{\epsilon}} \right)\, ,
\eeq
where $\omega,\bar{\omega}=t\pm x$ are coordinates on the boundary plane. The result matches the CFT$_2$ result upon the replacement $c=\bar{c}=3l/2G$. 

\fi

Note here that the screen was introduced only to regulate the length, not to define the geodesic in the first place.

\subsection{Covariant entanglement entropy in flat space}\label{sec:SEE}
The field theory calculations in the literature are based in the replica trick \cite{Bagchi:2014iea} or ``flat" extensions of the Cardy formula \cite{Jiang:2017ecm}. The vacuum result for a region consisting of an interval $[(0,0),(u,x)]$ in the  plane is given by
\beq\label{eq:SBMSFT}
S^{\text{BMS}_3}_{\text{EE}} = {{c_L}\over 6} \log {{x}\over{\epsilon}} +{{c_M}\over 6} {{u}\over{x}}\, ,
\eeq
where $u$ is the time-like coordinate while $x$ is space-like, and $c_L$, $c_M$ are the central charges of the theory. As before, $\epsilon$ is a regulator of the UV divergences. For the case of theories dual to Einstein gravity in asymptotically flat geometries, the central charges are $c_L=0$ and $c_M=3/G$. The result for entanglement entropy in this case does not diverge as we take the UV regulator to zero. This seems to imply that in a continuum QFT with BMS symmetry there are no UV correlations at arbitrarily small scales, which makes it possible to split the Hilbert space along some dividing surface. Clearly this signals a fundamental difference between two-dimensional conformal field theories and field theories with BMS$_3$ symmetry. 

In this section we follow the logic of  HRT \cite{Hubeny:2007xt} to propose a covariant approach for the computation of holographic entanglement entropy in the context of flat 2+1 dimensional Minkowski space.

As in previous sections, we work with Eddingston-Fikelstein coordinates and the line element of equation \eqref{eq:LineElement}. The boundary region ${\cal A}$ is located at the null plane $r\rightarrow\infty$ with $u$ and $\phi$ fixed. The boundary endpoints of the region ${\cal A}$ are dubbed $\partial_i{\cal A}$, and they are placed at $(u^{\partial}_i,\phi^{\partial}_i)$. In these coordinates, we hope to reproduce the BMSFT result for entanglement entropy as written in \eqref{eq:SBMSFT}.

We define a curve ${\cal S}$ with the following constraints 
\beq
\varphi_1=u-U(r)\, ,\quad \varphi_2 =\phi-\Phi(r)\, .
\eeq
The null vectors in equation \eqref{eq:NullVectorsN} are then specified by
\beq
\mu_{\pm}={{-r^2 \Phi'(1+U')}\over{1+r^2 \Phi'^2}}\pm {{r\sqrt{r^2 \Phi'^2-U'(2+U')}}\over{1+r^2 \Phi'^2}}\, ,  \quad {\cal N}=\sqrt{{{1+r^2 \Phi'^2}\over{2r^2\Phi'^2-2U'(2+U')}}}\, .
\eeq
The null expansions read
\beq\label{eq:diffs}
\begin{split}
\theta_+-\theta_-&\sim \left(  2\Phi'+r^2 \Phi'^3+r \Phi'' \right)\, , \\
\theta_++\theta_-&\sim \left(  r \Phi'(1+U')(\Phi'+r\Phi'')-(1+r^2 \Phi'^2)U''  \right)\, .
\end{split}
\eeq
The solutions with $\theta_{\pm}=0$ specify extremal curves. The second differential equation in \eqref{eq:diffs} can be solved by replacing
\beq
U(r)=c_1 -r+r\cos\left( \Phi(r)-c_2\right)\, ,
\eeq
which makes the second differential equation in \eqref{eq:diffs} proportional to the first. The constants $c_1$ and $c_2$ can be fixed by imposing that the curve ${\cal S}$ starts at ${\partial_1 {\cal A}}$. This implies $U(r\rightarrow \infty)=u^{\partial}_1$ and $\Phi(r\rightarrow\infty)=\phi^{\partial}_1$, which leads to $c_1=u^{\partial}_1$ and $c_2=\phi^{\partial}_1$. We thus have
\beq\label{eq:G}
U(r)=u^{\partial}_1 -r+r\cos\left(\Phi(r)-\phi^{\partial}_1 \right)\, .
\eeq
It is important to note that the solution \eqref{eq:G} matches the equation for the null plane $N_1$ in formula \eqref{eq:originalNi}. In the work \cite{Jiang:2017ecm}, this plane corresponds to the Cauchy horizon of a flat space cosmological solution that arises through the bulk Rindler map used to compute entanglement entropy. Here, the same plane arises naturally when studying extremal curves in flat space. 

The solution for $\Phi(r)$ can be obtained by solving the first differential equation in equation \eqref{eq:diffs}. The solution reads
\beq\label{eq:Fd1}
\Phi(r)=\pm \arctan{{{d_1}\over{\sqrt{r^2-d_1^2}}}}+d_2\, ,
\eeq
where $d_i$ are constants of integration. Imposing $\Phi(r\rightarrow \infty )=\phi^{\partial}_1$ implies $d_2=\phi^{\partial}_1$. The resulting curve specified by $\Phi(r)$ still depends on a constant of integration $d_1$, and corresponds to a null straight line along the null direction of the plane $N_1$ specified in \eqref{eq:G}. Solutions with different $d_1$ correspond to inequivalent null parallel lines and $d_1$ is the impact parameter or point of closest approach to our chosen origin. 

These parallel lines all go to the same point on the conformal boundary. However,
the Minkowski metric at the $r\rightarrow \infty$ boundary reads 
\beq
ds^2\rightarrow -du^2+r^2d\phi^2\, ,
\eeq
and so parallel null lines points with different $\phi$ will be distinguishable unless $r \Delta\phi =0$. The choice of $d_1$ is an ambiguity in associating null rays to points on future null infinity which must be fixed in a consistent way. 
A choice for $d_1$ can be mapped into another by changing the origin of our coordinate system. We will fix $d_1$ by setting it to zero, so that all the null rays we consider pass through the origin of our coordinate system. This choice has the nice feature that in our coordinate system, the null rays meet the natural cut-off surface at fixed $r$ at a right angle. They obey $r \Phi'(r)=0$ at large $r$. In other words, this is the requirement that the boundary point reached by our null ray should be stable under changes to our cut-off surface.

The solution for the extremal curve falling from $\partial_1 {\cal A}$ is then
\beq
U(r)=u^{\partial}_1\, , \quad \Phi(r)=\phi^{\partial}_1\, ,
\eeq
which corresponds to a null straight line in the radial direction. Unlike the curve studied in AdS/CFT, an extremal curve  in flat space cannot connect two different boundary points  $\partial_1 {\cal A}$  and  $\partial_2 {\cal A}$.   

That the points on the asymptotic boundary can not be connected could be circumvented by introducing a cut-off surface at a large fixed $r=\epsilon^{-1}$ and connecting the points on the regulated surface. 
 However, this geodesic will not be well defined in the limit where we remove the cut-off. In AdS, the extension of a geodesic connecting points on the regulated surface will reach the asymptotic boundary near the correct location. In flat space, the geodesic connecting these regulated boundary points will generically be spacelike and run off to the spacelike asymptotic boundary rather than future null infinity. In any case, the length of this regulated geodesic diverges in the $\epsilon\rightarrow0$ limit and there are no BMS$_3$ invariant counter-terms which can be added to the length so as to reproduce the correct entanglement entropy. In the following we will introduce a prescription motivated by the construction of \cite{Jiang:2017ecm} which does correctly reproduce the entanglement entropy.

So far, we have concluded that the extremal curves falling from the boundary correspond to radial null geodesic lines which we will refer to as $\gamma_i$. In general these curves do not intersect, so in order to connect the boundary points we need to introduce a third curve connecting $\gamma_1$ and $\gamma_2$. We choose points $y_i\in\gamma_i$ and connect them with a curve ${\cal S}'$. Imposing extremality of ${\cal S}'$ as above implies that ${\cal S}'$ is a straight line connecting $y_1$ with $y_2$. The total length functional is then
\beq
L_{\text{total}}=L^{\text{extr}}(\partial_1 {\cal A}, y_1)+L^{\text{extr}}(y_1,y_2)+L^{\text{extr}}(y_2 ,\partial {\cal A}_2 )= L^{\text{extr}}(y_1,y_2)\, ,
\eeq
where $L^{\text{extr}}(P,Q)$ is the length of a straight line from point $P$ to point $Q$, and we have used the fact that null geodesics have no length. The final step is to extremize the total length functional with respect to the locations of  $y_i$ along the null lines $\gamma_i$. We start with the following expression for the length between the points at the null geodesics
\beq\label{eq:auxLextr}
L^{\text{extr}}(y_1,y_2)=\sqrt{r_1^2+r_2^2-2r_1 r_2 \cos\left(\phi^{\partial}_1-\phi^{\partial}_2\right)-\left(u^{\partial}_2+r_2-u^{\partial}_1-r_1\right)^2}\, ,
\eeq
where $r_i$ is the radial location of the points $y_i$ along $\gamma_i$.  Solving for $\partial L_{\text{total}}/{\partial r_i}=0$ implies 
\beq
r_1={{u^{\partial}_2-u^{\partial}_1}\over{1-\cos\left(\phi^{\partial}_1-\phi^{\partial}_2\right)}}\, , \quad r_2=-{{u^{\partial}_2-u^{\partial}_1}\over{1-\cos\left(\phi^{\partial}_1-\phi^{\partial}_2\right)}}\, .
\eeq
Note that this is the condition that the straight line connecting $y_1$ and $y_2$ must be orthogonal to the null rays $\gamma_i$:
\begin{align}
\frac{\partial L^\text{extr}(y_1,y_2) }{\partial r_1} 
= \frac{\partial y_1^\mu}{\partial r_1} \frac{\partial L^\text{extr}(y_1,y_2) }{\partial y_1^\mu} 
= n_1^\mu t_\mu =0 \,,
\end{align}
where $t^\mu$ is tangent to the line connecting $y_1$ and $y_2$ and $n_i^\mu$ is the null vector tangent to $\gamma_i$. The plane orthogonal to $\gamma_i$ is the $N_i$ introduced in \eqref{eq:originalNi}, which includes $n_i^\mu$ since it is null. Thus extremising over the two endpoints imposes the condition that the line connecting $y_1$ and $y_2$ lies along $N_1 \cap N_2$, which is sufficient to fix it.

Replacing these back in \eqref{eq:auxLextr} gives
\beq\label{eq:auxLextrsol}
L^{\text{extr}}(y_1,y_2)=\left|{{u^{\partial}_2-u^{\partial}_1}\over{\tan{{\phi^{\partial}_2-\phi^{\partial}_1}\over{2}}}} \right| \, .
\eeq
We conclude that
\beq
S_{EE}={1\over {4G}}L_{\text{total}}^{\text{extremal}}={1\over 4G}\left|{{u^{\partial}_2-u^{\partial}_1}\over{\tan{{\phi^{\partial}_2-\phi^{\partial}_1}\over{2}}}} \right|\, ,
\eeq
which is the same answer obtained in the BMSFT. 

Summarizing, we have successfully computed holographic entanglement as the length of an extremal geodesic path homologous to ${\cal A}$. However, we had to consider an extremal geodesic path with corners.\footnote{Note that corners do not result on a finite contribution to the length as can be seen by smoothing them out and taking the corner limit.} The following question then arises; If we were happy to introduce two corners, why not more? In particular, by adding one more corner we would find an extremal curve with no length connecting the two boundary points. One approach would be to use the curve with the least corners such that it has finite length. Instead, we will think of the null rays $\gamma_i$ associated to points on the future null boundary as regulators which do not contribute to the length and find the minimal length curve which connects the two null rays.  This latter view point will lead us to propose an extrapolate dictionary for flat space.

\begin{figure}[t!]
\centering
\begin{subfigure}[t]{0.48\textwidth}
        \centering

\tdplotsetmaincoords{60}{100}
\begin{tikzpicture}[scale=5,tdplot_main_coords]

    \def\x{1}

  \filldraw[
        draw=black,%
        fill=black!20,%
    ]          (0,1,0)
            -- (1,1,0)
            -- (1,1/2,1/2)
            -- (0,1/2,1/2)
            -- cycle;

  \draw[
        darkgreen,dashed,very thick%
    ]          (3/4,1,0)
            -- (3/4,1/2,1/2) node [pos=0.3,above,name=int] {{\textcolor{black}{$\gamma_1$}}};

    \filldraw[
        draw=black,%
        fill=black!20, bottom color=gray, top color=white%
    ]          (0,0,0)
            -- (1,0,0)
            -- (1,1,1)
            -- (0,1,1)
            -- cycle;

  \draw[
        darkgreen,dashed,very thick%
    ]          (1/4,0,0)
            -- (1/4,1/2,1/2) node [pos=0.2,below=2,name=int] {{\textcolor{black}{$\gamma_2$}}};

  \filldraw[
        draw=black,%
        fill=black!20,bottom color=gray, top color=white%
    ]          (0,0,1)
            -- (1,0,1)
            -- (1,1/2,1/2)
            -- (0,1/2,1/2)
            -- cycle;

  \filldraw[
        draw=black,%
        fill=black!20,bottom color=gray, top color=white%
    ]          (0,0,1)
            -- (1,0,1)
            -- (1,1/2,1/2)
            -- (0,1/2,1/2)
            -- cycle;

\draw (1,0,0) node[below] {$N_2$};
\draw (1,1,0) node[below] {$N_1$};

  \draw[
         darkgreen,dashed,very thick%
    ]          (3/4,0,1)
            -- (3/4,1/2,1/2) node [pos=0,left,name=O1] {\textcolor{black}{$x_1$}};
  \draw[
          darkgreen,dashed,very thick%
    ]          (1/4,1,1)
            -- (1/4,1/2,1/2) node [pos=0,right,name=O2] {\textcolor{black}{$x_2$}};

  \draw[
       black,very thick%
    ]          (0,1/2,1/2)
            -- (1,1/2,1/2);

  \draw[
         -dot-=0,blue,very thick%
    ]          (3/4,0,1)
            -- (3/4,1/4,3/4);

  \draw[
        -dot-=0, blue,very thick%
    ]          (1/4,1,1)
            -- (1/4,3/4,3/4);

  \draw[
       -dot2-=0, -dot2-=1,blue,very thick%
    ]         (3/4,1/4,3/4)
            -- (1/4,3/4,3/4)   node [pos=1,below=2,name=OB2] {\textcolor{black}{$y_2$}}  node [pos=0,below=2,name=OB1] {\textcolor{black}{$y_1$}};

\end{tikzpicture}

        \caption{ }
    \end{subfigure}%
\quad
\begin{subfigure}[t]{0.48\textwidth}
        \centering

\tdplotsetmaincoords{60}{100}
\begin{tikzpicture}[scale=5,tdplot_main_coords]

    \def\x{1}

  \filldraw[
        draw=black,%
        fill=black!20,bottom color=gray, top color=white%
    ]          (0,1,0)
            -- (1,1,0)
            -- (1,1/2,1/2)
            -- (0,1/2,1/2)
            -- cycle;

  \draw[
        darkgreen,dashed,very thick%
    ]          (3/4,1,0)
            -- (3/4,1/2,1/2) ;

    \filldraw[
        draw=black,%
        fill=black!20,bottom color=gray, top color=white%
    ]          (0,0,0)
            -- (1,0,0)
            -- (1,1,1)
            -- (0,1,1)
            -- cycle;

  \draw[
        darkgreen,dashed,very thick%
    ]          (1/4,0,0)
            -- (1/4,1/2,1/2) ;

  \filldraw[
        draw=black,%
        fill=black!20,bottom color=gray, top color=white%
    ]          (0,0,1)
            -- (1,0,1)
            -- (1,1/2,1/2)
            -- (0,1/2,1/2)
            -- cycle;

\draw (1,0,0) node[below] {\textcolor{white}{$N_2$}};
\draw (1,1,0) node[below] {\textcolor{white}{$N_1$}};

  \draw[
          darkgreen,dashed,very thick%
    ]          (3/4,0,1)
            -- (3/4,1/2,1/2) node [pos=0,left,name=O1] {\textcolor{white}{$x_1$}};
  \draw[
        darkgreen,dashed,very thick%
    ]          (1/4,1,1)
            -- (1/4,1/2,1/2) node [pos=0,right,name=O2] {\textcolor{white}{$x_2$}};

  \draw[
       black,very thick%
    ]          (0,1/2,1/2)
            -- (1,1/2,1/2);

  \draw[
        -dot-=0,blue,very thick%
    ]          (3/4,0,1)
            -- (3/4,2/4,2/4);

  \draw[
       -dot-=0, blue,very thick%
    ]          (1/4,1,1)
            -- (1/4,2/4,2/4);

  \draw[
       -dot2-=0, -dot2-=1,blue,very thick%
    ]         (1/4,2/4,2/4)
            -- (3/4,2/4,2/4) ;

\end{tikzpicture}


        \caption{ }
    \end{subfigure}
    \caption{a) Calculation of Entanglement Entropy before extremizing. The red boundary points $x_i$ are connected to the blue points $y_i$, which live at the null lines $\gamma_i$.  b) Geometrical picture after extremizing the total length. The bulk points live at the intersections $N_1\cap  N_2\cap \gamma_i$.}\label{fig:2pt}
\end{figure}
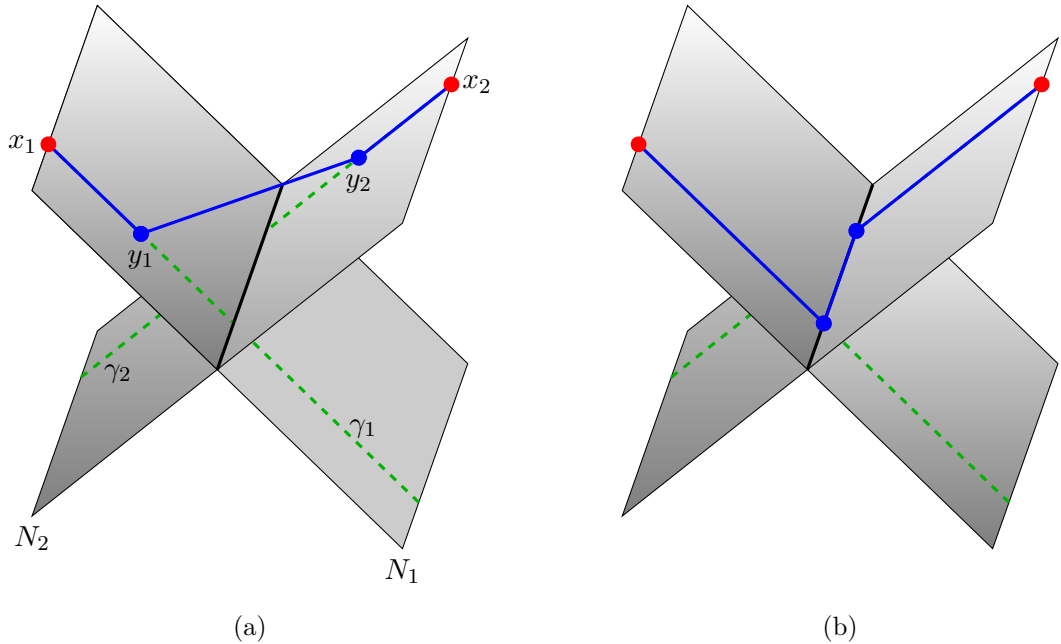 

\subsection{Relation to the replica trick}
We can now ask why this prescription works from the point of view of the replica trick. In \cite{Bagchi:2014iea}, The entanglement entropy in a generic BMS$_3$ field theory is computed by evaluating the two point function of twist fields $\Phi_n$ located at the boundary endpoints, and with quantum numbers
\beq
\Delta_n={{c_L}\over{24}}\left( 1-{1\over{n^2}} \right)\, ,\quad \xi_n={{c_M}\over{24}}\left( 1-{1\over{n^2}} \right)\, .
\eeq
In the case of semi-classical Einstein gravity in the bulk, we should set $c_L=0$ and take the limit of large $c_M$, which turns the twist field into a bulk propagating particle of mass $\xi_n$. The limit $n\rightarrow 1$ is equivalent to the probe limit
\beq
{{\xi_n}\over{c_M}} = 1-{1\over{n^2}}\ll 1\, .
\eeq
 Entanglement entropy is then computed as
\beq\label{eq:Replica}
S_{EE}=-\lim_{n\rightarrow 1}\partial_n \text{Tr}\rho_{{\cal A}}^n=-\lim_{n\rightarrow 1}\partial_n \langle \Phi_n(\partial_1 {\cal A})\Phi_n(\partial_2{\cal A})  \rangle\, .
\eeq
We must now propose a holographic prescription for computing this correlator.

In the standard AdS/CFT dictionary, the correlator in this expression is computed as the exponential of an on-shell action corresponding to a massive particle propagating along an extremal trajectory $X^{\mu}(s)$. The computation in the previous subsection leads us to propose that in flat space, the particles should be sourced at the null geodesics $\gamma_i$ associated to the boundary points in the correlator. In the probe limit, this means that the extremal trajectory $X^{\mu}(s)$ must be anchored along the $\gamma_i$ such that the endpoints are free to move along the $\gamma_i$ and should included in the extremisation.

The expression is
\beq\label{eq:PP}
\langle \Phi_n(\partial_1 {\cal A})\Phi_n(\partial_2{\cal A})  \rangle = e^{-m_n S_{\text{on-shell}}}\, ,
\eeq
where $m_n=\xi_n$ and
\beq
S_{\text{on-shell}}=\sqrt{\eta_{\mu\nu}\dot X^{\mu}\dot X^{\nu}}=L^{\text{extr}}_{\text{total}}\, .
\eeq
Plugging \eqref{eq:PP} in \eqref{eq:Replica}, we obtain the same formula used in the previous subsection
\beq
S_{EE}={1\over{4G}}L^{\text{extr}}_{\text{total}}\, .
\eeq
In sections \ref{sec:2pt} and \ref{sec:3pt}, we will use the same logic to calculate holographic BMS$_3$ two/three-point functions, and in section \ref{sec:block} we will consider Poincar\'e blocks.

\subsection{Holographic entanglement in flat space cosmologies (quotients)}\label{sec:FSC}
General classical solutions of Einstein gravity in three dimensions are locally equivalent. In the absence of a cosmological constant all solutions are locally flat, and are related to Minkowski space through quotients. An example of these solutions are Flat Space Cosmologies (FSC), whose metric reads
\beq
ds^2=M du^2-2du dr+J du d\phi+r^2d\phi^2\, .
\eeq
The dual BMSFT is expected to be a finite temperature theory with thermal cycle $(u,\phi)\sim(u+i\beta_u,\phi-i\beta_{\phi})$ with $\beta_u=\pi J M^{-3/2}$ and $\beta_{\phi}=2\pi M^{-1/2}$. The entanglement entropy in these field theories was computed in \cite{Basu:2015evh}, where they arrived to the following formula
\beq\label{eq:SEEFSC}
S_{EE}={1\over 4G} \left[\sqrt{M}\left(T+{J\over{2M}}L\right)\coth {{\sqrt{M}L}\over 2}-{J\over M} \right]\, .
\eeq
The same result can be achieved by extremizing the length of an extremal path homologous to ${\cal A}$ in the FSC geometry. Geodesics in this geometry are not straight lines, so the problem becomes a bit more complicated. The strategy we will follow here will be to compute the invariant length between two points in Minkowski space, and then map those points to points in the FSC using the coordinate transformation that implements the appropriate quotient. The coordinate transformation from Minkowski space to the FSC reads
\beq\label{eq:quotient}
\begin{split}
r&=\sqrt{M(t^2-x^2)+{{J^2}\over{4M}}}\, ,\\
\phi&=-{{1}\over{\sqrt{M}}}\log\left[ {{\sqrt{M}(t-x)}\over{r+{J\over{2\sqrt{M}}}}}\right]\, ,\\
u&={1\over M}\left( r-\sqrt{M}y-{J\over 2}\phi   \right)\, .
\end{split}
\eeq
The invariant length in Minkowski space reads
\beq\label{eq:MinkL}
\int ds =\sqrt{(x_1-x_2)^2+(y_1-y_2)^2-(t_1-t_2)^2}\, ,
\eeq
and under the coordinate transformation \eqref{eq:quotient}, the points $r_i,u_i,\phi_i$ are mapped to the points
\beq\label{eq:xyt}
\begin{split}
x_i&={{r_i}\over{\sqrt{M}}} \sinh \left( \sqrt{M}\phi_i \right)-{J\over{2M}}\cosh\left( \sqrt{M}\phi_i \right)\, , \\
y_i&=-{{r_i}\over{\sqrt{M}}}+{J\over{2\sqrt{M}}}\phi_i +u_i \sqrt{M}\, , \\
t_i&={{r_i}\over{\sqrt{M}}} \cosh \left( \sqrt{M}\phi_i \right)-{J\over{2M}}\sinh\left( \sqrt{M}\phi_i \right)\, .
\end{split}
\eeq
Replacing \eqref{eq:xyt} in \eqref{eq:MinkL} we obtain the length between two points in the FSC. The boundary points $\partial_i {\cal A}$ are located at $(u,\phi)=(u^{\partial}_i,\phi^{\partial}_i)$ in the future null plane. We need to find the extremal curve that starts at these boundary points. In the original minkowski coordinates the curve is a null line, and so the invariant length between any point in the extremal curve and the boundary point is zero. We thus need to find the points in the flat space cosmological solution that have zero invariant distance with the boundary points $\partial_i {\cal A}$.

 The distance between a bulk point $y_i$ and $\partial_i {\cal A}$ will be infinite unless the bulk point lies on a special hypersurface that we call $N^{\text{FSC}}_i$. 
\beq\label{eq:FSCNi}
N^{\text{FSC}}_i\, :\, 0=M(u_i-u^{\partial}_i)+{J\over 2}(\phi_i-\phi^{\partial}_i)+2r_i \sinh^2{{\sqrt{M}(\phi_i-\phi^{\partial}_i)}\over{2}}-{J\over {2\sqrt{M}}}\sinh \left( \sqrt{M}(\phi_i-\phi^{\partial}_i)\right)\, .
\eeq
The invariant length between a point $y_i\in N^{\text{FSC}}_i$ and  $\partial_i {\cal A}$ reads
\beq
L(y_i,\partial_i{\cal A})={1\over M}\left| J\sinh\left(  \sqrt{M}{{\phi_i - \phi^{\partial}_i}\over 2}  \right) +\sqrt{M} r_i \sinh\left( \sqrt{M} (\phi_i-\phi^{\partial}_i) \right) \right| \, ,
\eeq
and so null lines are geodesics of constant $\phi_i=\phi^{\partial}_i$. All in all, the null geodesic in the flat space cosmological solution reads
\beq
\phi_i=\phi^{\partial}_i\, , \quad u_i=u^{\partial}_i\, .
\eeq
Just like in the previous section, the extremal geodesic falling from $\partial_1 {\cal A}$ does not intersect the one falling from $\partial_2 {\cal A}$. This forces us to connect the null lines with an extra extremal curve connecting the two different lightlike ones.  We choose a bulk point $y_1\in \gamma_1$ and another point $y_2\in\gamma_2$, such that the total length is
\beq
L_{\text{total}}=L^{\text{extr}}(y_1,\partial_1{\cal A})+L^{\text{extr}}(y_1,y_2)+L^{\text{extr}}(y_2,\partial_2{\cal A})=L^{\text{extr}}(y_1,y_2)\, .
\eeq
Extremizing the total length with respect to the location of the points $y_i$ along the null lines involves a morally similar computation as the one in the previous section. The process is a bit more cumbersome as the length of the geodesics is more complicated as a function of $r_i$, but the final answer matches the result written in equation \eqref{eq:SEEFSC}.

\section{Two-point and Three-point BMS$_3$ correlation functions}
In this section we generalize the worldline methods used in section \ref{sec:SEE} to commute entanglement entropy and we apply them to the computation of two- and three-point correlation functions of BMS$_3$ operators in the cylinder. The correlators we are hoping to compute have been written above in equations \eqref{eq:lowpt}.

As a starting point we will consider primaries with $\Delta=1$ in the $1\ll \xi \ll c_M$ probe limit. In section \ref{sec:SPIN} we will consider the spinning case $\Delta\neq 1$. The resulting primaries should correspond to massive scalars, and so we expect the correlator to be computed with the exponential of a scalar on-shell action. The prescription is to compute a Poincar\'e invariant by connecting the null lines $\gamma_i$ falling from the boundary points with a geodesic network, and then extremizing the answer with respect to the free parameters of the problem. We first consider the two-point function. 

\subsection{Scalar two-point function}\label{sec:2pt}
In order to compute the two-point function we place two primaries of mass $\xi$ at the null boundary, with cylinder coordinates $x_i=(u^{\partial}_i,\phi^{\partial}_i)$, for $i=1,2$. We want to connect these two points by geodesics in such a way that the total on-shell action is extremal. The first step is to connect each boundary point with a bulk point $y_i=(r_i,u_i,\phi_i)$ located at the  corresponding null geodesic $\gamma_i$. The bulk points are then connected with a space-like geodesic, giving rise to a total on-shell action
\beq
S_{\text{total}}=\xi \left[L(y_1,y_2)+\sum_{i=1,2} L(x_i,y_i) \right]= L(y_1,y_2) \, . 
\eeq
The extremization of this function with respect to the locations of $y_i$ follows exactly the same logic as the one between formulas \eqref{eq:auxLextr} and \eqref{eq:auxLextrsol}. The solution for the extremum is $y_i=N_1\cap N_2\cap \gamma_i$. We thus conclude 
\beq\label{eq:Lmin2pt}
S^{\text{extr}}_{\text{total}}=\xi \left|{{u^{\partial}_2-u^{\partial}_1}\over{\tan{{\phi^{\partial}_2-\phi^{\partial}_1}\over{2}}}} \right| \, .
\eeq
We recover the result for the scalar two point function by exponentiating the extremal length
\beq
\langle \Phi_{0,\xi} \Phi_{0,\xi}\rangle =e^{-S^{\text{extr}}_{\text{total}}}=e^{{-\xi{{u^{\partial}_2-u^{\partial}_1}\over{\tan{{\phi^{\partial}_2-\phi^{\partial}_1}\over 2}}}}}\, .
\eeq
The calculation in this section is exactly the same as the calculation of holographic entanglement entropy. We next study a slightly more complicated case; The scalar three-point function.

\subsection{Scalar three-point function}\label{sec:3pt}
\begin{figure}[t!]
\centering
\begin{subfigure}[t]{0.48\textwidth}
        \centering

\tdplotsetmaincoords{60}{100}
\begin{tikzpicture}[scale=6,tdplot_main_coords]

    \def\x{1}

    \filldraw[
        draw=black,%
        fill=black!20,bottom color=gray, top color=white%
    ]          (0,1/2,0)
            -- (1,1/2,0)
            -- (1,1,1/2)
            -- (0,1,1/2)
            -- cycle;

  \filldraw[
        draw=black,%
        fill=black!20,bottom color=gray, top color=white%
    ]          (0,0,1/2)
            -- (1,0,1/2)
            -- (1,1/2,0)
            -- (0,1/2,0)
            -- cycle;

  \draw[
         darkgreen,dashed,very thick%
    ]          (3/4,0,1/2)
            -- (3/4,1/2,0) node [pos=0,left,name=O1] {\textcolor{black}{$x_1$}};
  \draw[
          darkgreen,dashed,very thick%
    ]          (1/4,1,1/2)
            -- (1/4,1/2,0) node [pos=0,right,name=O2] {\textcolor{black}{$x_2$}};

  \draw[
       black,very thick%
    ]          (0,1/2,0)
            -- (1,1/2,0)  node [pos=1,below,name=int] {{\small\textcolor{black}{$N_1\cap N_2$}}};

  \filldraw[
        draw=black,%
        fill=black!20,bottom color=gray, top color=white%
    ]          (1/2,1/2,0)
            -- (0,1,1/2)
            -- (0,0,1/2)
            -- cycle;

  \draw[
          darkgreen,dashed,very thick%
    ]          (0,1/4,1/2)
            -- (1/4,1/4,1/4) node [pos=0,above,name=O3] {\textcolor{black}{$x_3$}};

  \draw[
       black,very thick%
    ]          (1/2,1/2,0)
            -- (0,1,1/2)  node [pos=1,above,name=int] {{\small\textcolor{black}{$N_2\cap N_3$}}};

  \draw[
       black,very thick%
    ]          (1/2,1/2,0)
            -- (0,0,1/2)  node [pos=1,above,name=int] {{\small\textcolor{black}{$N_1\cap N_3$}}};

  \draw[
         -dot-=0,blue,very thick%
    ]          (3/4,0,1/2)
            -- (3/4,1/4,1/4);

  \draw[
         -dot-=0,blue,very thick%
    ]          (0,1/4,1/2)
            -- (1/8,1/4,{1/2-1/8});

  \draw[
         -dot-=0,blue,very thick%
    ]          (1/4,1,1/2)
            -- (1/4,7/8,{1/2-1/8});

  \draw[
         -dot2-=0,blue,very thick%
    ]          (3/4,1/4,1/4)
            -- (1/2+0.1,1/2,3/8)  node [pos=0,below,name=int] {{\small\textcolor{black}{$y_1$}}};

  \draw[
         -dot2-=0,blue,very thick%
    ]         (1/8,1/4,{1/2-1/8})
            --(1/2+0.1,1/2,3/8)   node [pos=0,right,name=int] {{\small\textcolor{black}{$y_3$}}};

  \draw[
         -dot2-=0, -dot3-=1,blue,very thick%
    ]         (1/4,7/8,{1/2-1/8})
            -- (1/2+0.1,1/2,3/8) node [pos=1,below,name=int] {{\small\textcolor{black}{$y_v$}}}  node [pos=0,below=2,right,name=int] {{\small\textcolor{black}{$y_2$}}};

\end{tikzpicture}

        \caption{ }
    \end{subfigure}%
\quad
\begin{subfigure}[t]{0.48\textwidth}
        \centering


\tdplotsetmaincoords{52}{100}
\begin{tikzpicture}[scale=4,tdplot_main_coords]

    \def\x{1}

    \filldraw[
        draw=black,%
        fill=black!20,left color=gray,right color=white%
    ]          (0,1/2,0)
            -- (1,1/2,0)
            -- (2,1,1/2)
            -- (0,1,1/2)
            -- cycle;

  \filldraw[
        draw=black,%
        fill=black!20,right color=gray,left color=white%
    ]          (0,0,1/2)
            -- (2,0,1/2)
            -- (1,1/2,0)
            -- (0,1/2,0)
            -- cycle;

  \draw[
         darkgreen,dashed,very thick%
    ]          (1+3/4,0,1/2)
            -- (1+3/4,1/8,1/2-1/8) node [pos=0,left,name=O1] {\textcolor{black}{$x_1$}};
  \draw[
          darkgreen,dashed,very thick%
    ]          (1/4,1,1/2)
            -- (1/4,1/2,0) node [pos=0,right,name=O2] {\textcolor{black}{$x_3$}};

  \draw[
       black,very thick%
    ]          (0,1/2,0)
            -- (1,1/2,0)  node [pos=0.75,right,name=int] {{\small\textcolor{black}{$N_1\cap N_3$}}};

  \filldraw[
        draw=black,%
        fill=black!20,bottom color=gray, top color=white%
    ]          (1/2,1/2,0)
            -- (0,1,1/2)
            -- (0,0,1/2)
            -- cycle;

  \filldraw[
        draw=black,%
        fill=black!20,bottom color=white, top color=gray%
    ]          (1,1/2,0)
            -- (2,0,1/2)
            -- (2,1,1/2)
            -- cycle;

  \draw[
          darkgreen,dashed,very thick%
    ]          (0,1/4,1/2)
            -- (1/4,1/4,1/4) node [pos=0,above,name=O3] {\textcolor{black}{$x_4$}};

  \draw[
          darkgreen,dashed,very thick%
    ]          (2,3/4,1/2)
            -- ({2-3/6},3/4,{1/2-(3/6)*1/2}) node [pos=0,below,name=O3] {\textcolor{black}{$x_2$}};

  \draw[
       black,very thick%
    ]          (1/2,1/2,0)
            -- (0,1,1/2)  node [pos=1,above,name=int] {{\small\textcolor{black}{$N_3\cap N_4$}}};

  \draw[
       black,very thick%
    ]          (1/2,1/2,0)
            -- (0,0,1/2)  node [pos=1,above,name=int] {{\small\textcolor{black}{$N_1\cap N_4$}}};

  \draw[
       black,very thick%
    ]          (1,1/2,0)
            -- (2,0,1/2)  node [pos=1,left,name=int] {{\small\textcolor{black}{$N_1\cap N_2$}}};

  \draw[
       black,very thick%
    ]          (1,1/2,0)
            -- (2,1,1/2)  node [pos=1,right,name=int] {{\small\textcolor{black}{$N_2\cap N_3$}}};

  \draw[
         -dot-=0,blue,very thick%
    ]          (2-1/4,0,1/2)
            --  (2-1/4,1/16,1/2-1/16);

  \draw[
         -dot-=0,blue,very thick%
    ]          (2,3/4,1/2)
            -- (2-2/8,3/4,1/2-2/16);

  \draw[
         -dot-=0,blue,very thick%
    ]          (0,1/4,1/2)
            -- (0+1/8,1/4,1/2-1/8);

  \draw[
         -dot-=0,blue,very thick%
    ]          (1/4,1,1/2)
            -- (1/4,1-1/8,1/2-1/8);

  \draw[
         -dot2-=0,blue,very thick%
    ]         (2-1/4,1/16,1/2-1/16)
            -- (1/2+1/2+0.1,1/2,1/4)   node [pos=0,right=3,above=2,name=int] {{\small\textcolor{black}{$y_1$}}};

  \draw[
         -dot2-=0,blue,very thick%
    ]          (2-2/8,3/4,1/2-2/16)
            -- (1/2+1/2+0.1,1/2,1/4)   node [pos=0,left,name=int] {{\small\textcolor{black}{$y_2$}}};

  \draw[
         -dot2-=0,blue,very thick%
    ]        (0+1/8,1/4,1/2-1/8)
            -- (1/2+0.1,1/2,1/4)   node [pos=0,right,name=int] {{\small\textcolor{black}{$y_4$}}};

  \draw[
         -dot2-=0, -dot3-=1,blue,very thick%
    ]         (1/4,1-1/8,1/2-1/8)
            -- (1/2+0.1,1/2,1/4) node [pos=0,below,name=int] {{\small\textcolor{black}{$y_3$}}};

  \draw[
         -dot3-=0, -dot3-=1,blue,very thick%
    ]        (1/2+0.1,1/2,1/4)
            -- (1/2+1/2+0.1,1/2,1/4) node [pos=1,left=4,above,name=int] {{\small\textcolor{black}{$y_{v}$}}}  node [pos=0,right=5,above=3,name=int] {{\small\textcolor{black}{$y_{v'}$}}};

\end{tikzpicture}


        \caption{ }
    \end{subfigure}
    \caption{a) Set-up for the calculation of the holographic three point function before extremizing. The red boundary points $x_i$ are connected to the blue points $y_i$, which live at the null lines $\gamma_i$.  The bulk points are then connected to a completely arbitrary vertex point $y_v$, drawn in purple.  b) Set-up for the calculation of the holographic scalar global block before extremizing. The red boundary points $x_i$ are connected to the blue points $y_i$, which live at the null lines $\gamma_i$. The bulk points are then connected to two completely arbitrary vertex points $y_v$ and $y_{v'}$, drawn in purple. These two vertex are connected with a line where a particle with mass $\xi_p$ is propagating. }\label{fig:3pt}
\end{figure}
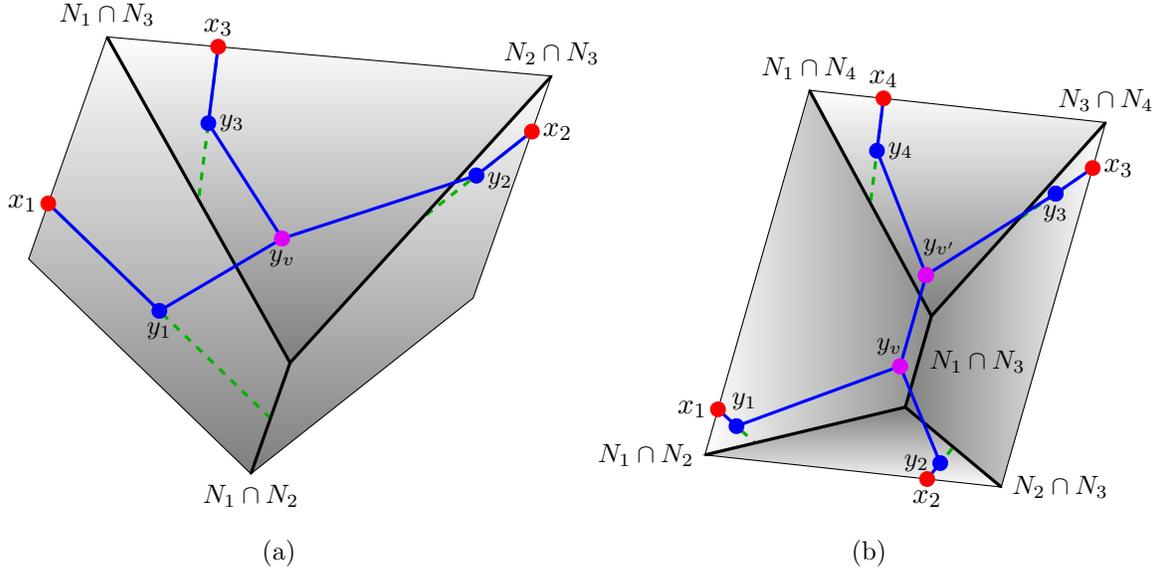

The prescription is the following. There are three primaries located at the points $x_k=(u^{\partial}_k,\phi^{\partial}_k)$, with $k=1,2,3$, and mass $\xi_k$. We want to connect the boundary points with geodesics in such a way that we obtain an extremum of the total length. The extremal curves falling from the boundary are null lines $\gamma_k$.  We thus connect each null line with a common bulk vertex. For this we choose three points $y_k \in \gamma_k$ and connect them to the vertex point $y_{v}=(r_v,u_v,\phi_v)$. A picture of the set-up can be found in figure \ref{fig:3pt}.a). The total length weighted by the corresponding masses is then
\beq
S_{\text{total}}=\sum_{k=1,2,3} \xi_k \left[ L(x_k,y_k) + L(y_k,y_v)\right] =\sum_{k=1,2,3} \xi_k  L(y_k,y_v)\, . 
\eeq
 The  free parameters are the radial locations of $y_k$ along $\gamma_k$, which we denote by $r_k$, and the vertex point $y_{v}$. The action as a function of these variables reads
\beq
S_{\text{total}}=\sum_{k=1,2,3} \xi_k \left[ \sqrt{ 
r_k\left(u_v-u^{\partial}_k +2r_v\sin^2{{\phi_v-\phi^{\partial}_k}\over 2} \right)-(u_v-u^{\partial}_k)(2r_v+u_v-u^{\partial}_k)
}\right] \, .
\eeq
Extremizing with respect to $r_k$ gives rise to three equations imposing the orthogonality between each of the lines $y_k$-$y_v$ and $\gamma_k$. Namely
\beq
u_v-u^{\partial}_i+2r_v \sin^2{{\phi_v-\phi^{\partial}_i}\over 2} =0\,, \quad \forall i=1,2,3 \, .
\eeq
Solving these conditions fixes the position of $y_v$ completely. Note that each of these equations corresponds to the planes $N_i$ of equation \eqref{eq:originalNi}. The point $y_v$ must live at $N_1\cap N_2\cap N_3$.  The result can be written simply in Minkowski coordinates
\beq\label{eq:xv}
\begin{split}
u_v+r_v&={{u^{\partial}_3\sin\left(\phi^{\partial}_1-\phi^{\partial}_2\right)-u^{\partial}_2\sin\left(\phi^{\partial}_1-\phi^{\partial}_3\right)+u^{\partial}_1\sin\left(\phi^{\partial}_2-\phi^{\partial}_3\right)}\over{\sin\left(\phi^{\partial}_1-\phi^{\partial}_2\right)-\sin\left(\phi^{\partial}_1-\phi^{\partial}_3\right)+\sin\left(\phi^{\partial}_2-\phi^{\partial}_3\right)}}\, ,\\
r_v \cos\phi_v&={{(u^{\partial}_3-u^{\partial}_2)\sin\phi^{\partial}_1-(u^{\partial}_3-u^{\partial}_1)\sin\phi^{\partial}_2+(u^{\partial}_2-u^{\partial}_1)\sin\phi^{\partial}_3}\over{\sin\left(\phi^{\partial}_1-\phi^{\partial}_2\right)-\sin\left(\phi^{\partial}_1-\phi^{\partial}_3\right)+\sin\left(\phi^{\partial}_2-\phi^{\partial}_3\right)}}\, ,\\
r_v \sin\phi_v&=-{{(u^{\partial}_3-u^{\partial}_2)\cos\phi^{\partial}_1-(u^{\partial}_3-u^{\partial}_1)\cos\phi^{\partial}_2+(u^{\partial}_2-u^{\partial}_1)\cos\phi^{\partial}_3}\over{\sin\left(\phi^{\partial}_1-\phi^{\partial}_2\right)-\sin\left(\phi^{\partial}_1-\phi^{\partial}_3\right)+\sin\left(\phi^{\partial}_2-\phi^{\partial}_3\right)}}\, .
\end{split}
\eeq
Replacing these back in $S_{\text{total}}$ results in
\beq\label{eq:li3}
S^{\text{extr}}_{\text{total}}=\sum_{k=1,2,3} \xi_k  {{\sum_{i<j}(-1)^{1+i+j}(u^{\partial}_i-u^{\partial}_j)\cos(\phi^{\partial}_k-\phi^{\partial}_i)}
\over
{\sum_{i<j}(-1)^{1+i+j}\sin(\phi^{\partial}_i-\phi^{\partial}_j)}} \, .
\eeq
We recover the desired result for the scalar three-point function after exponentiating the extremal value of the length.
\beq
\langle \Phi_{0,\xi_1} \Phi_{0,\xi_2}\Phi_{0,\xi_3}\rangle =e^{-S^{\text{extr}}_{\text{total}}}=\prod_k{e^{
\xi_k{{\sum_{i<j}(-1)^{1+i+j}(u^{\partial}_i-u^{\partial}_j)\cos(\phi^{\partial}_k-\phi^{\partial}_i)}
\over
{\sum_{i<j}(-1)^{1+i+j}\sin(\phi^{\partial}_i-\phi^{\partial}_j)}}
} }\, .
\eeq
Note that even though we did not need to solve for $r_k$ in order to compute the value of $S_{\text{total}}$ at the extremum, the extremality condition for the components of $y_v$ fixes them. Performing this calculation yields
\beq
y_k={{\gamma_k\cap N_i+\gamma_k\cap N_j}\over 2}\, , \quad \text{with}\, \quad k\neq i\neq j\, .
\eeq

\section{Holographic scalar global  BMS$_3$ blocks (Poincar\'e blocks)}\label{sec:block}
We are now ready to propose a calculation of a Poincar\'e invariant object that computes a global BMS$_3$ block holographically in the probe limit. The strategy is very similar to the one used for computing low-point correlation functions and it emulates the holographic construction of global blocks in the context of AdS/CFT. 

The result we are hoping to reproduce in this section corresponds to the scalar part of equation \eqref{eq:Block}. As we will see, the exponential involving $\xi$ and $\xi_p$ can be obtained holographically with the technology described in this note so far. The terms involving the spin $\Delta$ and $\Delta_p$ require additional tools, and we will discuss how to obtain the holographic result in section \ref{sec:SPIN}.

The prescription is to extremize the length of a geodesic network weighted by the mass of the corresponding primaries. We first introduce four bulk points $y_{k}$. Each of these must live on the corresponding null geodesic $\gamma_k$ such that the curve connecting $y_{k}$ and the boundary points $x_{k}$  is extremal. The next step is to introduce two vertex points $y_v$ and  $y_{v'}$. Two geodesics join the points $y_{k=1,2}$ with $y_v$, while two different geodesics join the points $y_{k=3,4}$ with $y_{v'}$. All the geodesics we have introduced until now are weighted by the mass $\xi$ of the external primaries of the block. The last step is to join the vertex points $y_{v}$ and $y_{v'}$ with a geodesic whose length is weighted by the mass of the exchanged family, $\xi_p$. A figure of this set-up can be found in \ref{fig:3pt}.b).

The total length we have to extremize reads
\beq
S_{\text{total}}=\xi \left[\sum^{4}_{k=1} L(x_k,y_k)+\sum^2_{k=1}L(y_k,y_{v}) +\sum^4_{k=3}L(y_k,y_{v'}) \right] +\xi_p L(y_{v'},y_{v})\, .
\eeq
The free parameters with respect to which $S_{\text{total}}$ must be extremal are the bulk vertex points $y_v$ and $y_{v'}$, and the radial locations of $y_k$ along $\gamma_k$, which we dub $r_k$. The action in terms of these variables reads
\beq
\begin{split}
S_{\text{total}}&=\xi \sum_{k=1,2} \left[ \sqrt{ 
2r_k\left(u_v-u^{\partial}_k +2r_v\sin^2{{\phi_v-\phi^{\partial}_k}\over 2} \right)-(u_v-u^{\partial}_k)(2r_v+u_v-u^{\partial}_k)
}\right] \\
&+\xi \sum_{k=3,4} \left[\sqrt{ 
2r_k\left(u_{v'}-u^{\partial}_k +2r_{v'}\sin^2{{\phi_{v'}-\phi^{\partial}_k}\over 2} \right)-(u_{v'}-u^{\partial}_k)(2r_{v'}+u_{v'}-u^{\partial}_k)
}\right] \\
&+\xi_p \sqrt{r_v^2+r_{v'}^2-2r_v r_{v'}\cos\left( \phi_v-\phi_{v'}\right)-(u_v+r_v-u_{v'}-r_{v'})^2}\, .
\end{split}
\eeq
We now solve for $\partial S_{\text{total}}/\partial r_k=0$. These equations read
\beq\label{eq:vvp}
\begin{split}
u_v-u^{\partial}_1+2r_v \sin^2 {{\phi_v-\phi^{\partial}_1}\over 2}&=0\, ,\\
u_v-u^{\partial}_2+2r_v \sin^2 {{\phi_v-\phi^{\partial}_2}\over 2}&=0\, ,\\
u_{v'}-u^{\partial}_3+2r_{v'} \sin^2 {{\phi_{v'}-\phi^{\partial}_3}\over 2}&=0\, ,\\
u_{v'}-u^{\partial}_4+2r_{v'} \sin^2 {{\phi_{v'}-\phi^{\partial}_4}\over 2}&=0\, .
\end{split}
\eeq
As before, we can think of these as orthogonality constraints which constrain the location of the bulk vertices. They imply that $y_v$ lives somewhere on $N_1\cap N_2$, which is the minimal spacelike line connecting the null rays which appeared in the computation of the two-point function. Similarly, $y_{v'}$ lives somewhere on  $N_3\cap N_4$. Using \eqref{eq:vvp} to simplify $S_{\text{total}}$ yields
\beq\label{eq:trick}
\begin{split}
S_{\text{total}}&=\xi \left[ \sum_{k=1,2}\sqrt{\left( r_v \sin\left( \phi_v-\phi^{\partial}_k\right) \right)^2} +\sum_{k=3,4}\sqrt{\left( r_{v'} \sin\left( \phi_{v'}-\phi^{\partial}_k\right) \right)^2} \right]\\
&+\xi_p\sqrt{
r_v^2+r_{v'}^2-2r_v r_{v'}\cos(\phi_v-\phi_{v'})-(u_v+r_v-u_{v'}-r_{v'})^2
}\, .
\end{split}
\eeq

Expression \eqref{eq:trick} justifies introducing the parameters
\beq\label{eq:vvp2}
\begin{split}
s_k&=r_v \sin\left( \phi_v-\phi^{\partial}_k\right)\quad \text{for}\quad k=1,2\, ,\\
s'_k&= r_{v'} \sin\left( \phi_{v'}-\phi^{\partial}_k\right)\quad \text{for}\quad k=3,4 \, ,
\end{split}
\eeq
which correspond to the lengths on either side of $y_v$ along $N_1 \cap N_2$ and similarly for $y_{v'}$:
\beq\label{eq:sols1s3}
s_1 + s_2 =(u^{\partial}_2-u^{\partial}_1)\cot{{\phi^{\partial}_1-\phi^{\partial}_2}\over 2}\, , \quad s'_3 + s'_4=(u^{\partial}_4-u^{\partial}_3)\cot{{\phi^{\partial}_3-\phi^{\partial}_4}\over 2}\, .
\eeq

Using this parametrisation, the location of the vertices can be written in terms of $s_2$ and $s'_4$.
The position of $y_v$ reads
\beq\label{eq:solv}
\begin{split}
t_v&=u_v+r_v={{u^{\partial}_1-u^{\partial}_2 \cos(\phi^{\partial}_1-\phi^{\partial}_2)+s_2\sin(\phi^{\partial}_1-\phi^{\partial}_2)}\over{1-\cos(\phi^{\partial}_1-\phi^{\partial}_2)}}\, , \\
x_v&=r_v \cos\phi_v={{(u^{\partial}_1-u^{\partial}_2) \cos(\phi^{\partial}_2)+s_2(\sin(\phi^{\partial}_1)-\sin(\phi^{\partial}_2))}\over{1-\cos(\phi^{\partial}_1-\phi^{\partial}_2)}}\, , \\
y_v&=r_v \sin\phi_v={{(u^{\partial}_1-u^{\partial}_2) \sin(\phi^{\partial}_2)+s_2(\cos(\phi^{\partial}_2)-\cos(\phi^{\partial}_2))}\over{1-\cos(\phi^{\partial}_1-\phi^{\partial}_2)}}\, .
\end{split}
\eeq
Similarly, for $y_{v'}$ we have
\beq\label{eq:solvp}
\begin{split}
t_{v'}&=u_{v'}+r_{v'}={{u^{\partial}_3-u^{\partial}_4 \cos(\phi^{\partial}_3-\phi^{\partial}_4)+s'_4\sin(\phi^{\partial}_3-\phi^{\partial}_4)}\over{1-\cos(\phi^{\partial}_3-\phi^{\partial}_4)}}\, , \\
x_{v'}&=r_{v'} \cos\phi_{v'}={{(u^{\partial}_3-u^{\partial}_4) \cos(\phi^{\partial}_4)+s'_4(\sin(\phi^{\partial}_3)-\sin(\phi^{\partial}_4))}\over{1-\cos(\phi^{\partial}_3-\phi^{\partial}_4)}}\, , \\
y_{v'}&=r_{v'} \sin\phi_{v'}={{(u^{\partial}_3-u^{\partial}_4) \sin(\phi^{\partial}_3)+s'_4(\cos(\phi^{\partial}_4)-\cos(\phi^{\partial}_4))}\over{1-\cos(\phi^{\partial}_3-\phi^{\partial}_4)}}\, .
\end{split}
\eeq

We can now replace formulas \eqref{eq:solv}, \eqref{eq:solvp}, and \eqref{eq:sols1s3} in $S_{\text{total}}$. The resulting on-shell action now depends on the parameters $s_2$, and $s'_4$.
\beq
\begin{split}
S_{\text{total}}&=\xi \left[\sqrt{\left( (u^{\partial}_2-u^{\partial}_1)\cot{{\phi^{\partial}_1-\phi^{\partial}_2}\over 2}\right)^2}+\sqrt{\left( (u^{\partial}_4-u^{\partial}_3)\cot{{\phi^{\partial}_3-\phi^{\partial}_4}\over 2}\right)^2}
 \right]\\
&+\xi_p\sqrt{
F(s_2,s'_4)
}\, .
\end{split}
\eeq
We are now ready to extremize with respect to $s_2$ and $s'_4$. This means we have to solve for $\partial S_{\text{total}}/\partial s_2=\partial S_{\text{total}}/\partial s'_4=0$, which fully determine $s_2$ and $s'_4$.\footnote{The 6 coordinates of $y_v$ and $y_{v'}$ have been parametrised by 2 variables $s_2$ and $s'_4$, hiding 4 extremality constraints. These can be used to fix the 4 $r_k$ along the external legs. However, these are not needed to compute the value of the action at the extremum so we have omitted them from the discussion.}
After replacing the solution back in $S_{\text{total}}$, we obtain
\beq
\begin{split}
S_{\text{total}}&=\xi \left[ (u^{\partial}_2-u^{\partial}_1)\cot{{\phi^{\partial}_2-\phi^{\partial}_1}\over 2}+ (u^{\partial}_4-u^{\partial}_3)\cot{{\phi^{\partial}_4-\phi^{\partial}_3}\over 2}
 \right]\\
&+\xi_p
{{u^{\partial}_{43}\sin\phi^{\partial}_{21}+u^{\partial}_{42}\sin\phi^{\partial}_{13}+u^{\partial}_{14}\sin\phi^{\partial}_{23}+u^{\partial}_{23}\sin\phi^{\partial}_{14}+u^{\partial}_{31}\sin\phi^{\partial}_{24}+u^{\partial}_{12}\sin\phi^{\partial}_{34}}
\over
{
8\sin{{\phi^{\partial}_{34}}\over 2}\sin{{\phi^{\partial}_{12}}\over 2}\sqrt{\sin{{\phi^{\partial}_{13}}\over 2}\sin{{\phi^{\partial}_{14}}\over 2}\sin{{\phi^{\partial}_{23}}\over 2}\sin{{\phi^{\partial}_{24}}\over 2}}
}}
\, .
\end{split}
\eeq

This is the answer we expect from the BMSFT result in \cite{Bagchi:2017cpu}. This can be seen by writing everything in terms of cross-ratios and coordinates in the plane
\beq\label{eq:Result4PtFunction}
S_{\text{total}}=\left({{t_{ij}}\over{x_{ij}}}\sum_k{{\xi_{ijk}}\over3}\right) 
+\left(   {{t}\over{1-x}} {{\xi_{231}+\xi_{234}}\over 3} -3{{t}\over{x}}(\xi_{341}+\xi_{342})\right)
+\left( 2\xi{t\over x}-\xi_p {{t}\over{x\sqrt{1-x}}}\right).
\eeq
Here the first two parenthesis match the prefactors in \eqref{eq:4PtFunction}, while the third term corresponds to the contribution from $g_{p}(x,t)$.

\section{Spinning correlators and Poincar\'e blocks}\label{sec:SPIN}
In the previous sections we have studied global blocks concerning primaries with vanishing $L_0$ eigenvalue. The representations of the BMS$_3$ algebra are usually labeled by two quantum numbers $\Delta$ and $\xi$, which are the eigenvalues of $L_0$ and $M_0$ respectively acting on the highest weight state. 
In this section, we compute low-point functions and global blocks involving general representations with non-vanishing $\Delta$ and $\xi$. Note that these representations of the BMS$_3$ algebra are not unitary \cite{Barnich:2014kra,Campoleoni:2016vsh}. It would also be interesting to understand the holography of blocks involving unitary irreducible representations of the BMS$_3$ algebra. However, to our knowledge there are no results concerning these blocks in the literature. 

In the $\Delta=0$ case, the parameter $\xi$ was associated to the mass of a particle propagating in flat space. We thus expect that the parameter $\Delta$ is associated with the spin of such particle. In \cite{Castro:2014tta}, an action of a massive spinning particle (anyon) was used to compute entanglement entropy in the context of AdS/CFT in a theory of topologically massive gravity. We will use the same action in this work. 
\beq\label{eq:Sspin}
S=\int d\sigma\left[ \xi \sqrt{\eta_{\mu \nu}\dot{X}^{\mu}\dot{X}^{\nu}} +\Delta \left( \tilde{n}\cdot \nabla n \right) \right]+S_{\text{constraints}}\, ,
\eeq
where $\tilde n$ is a space-like vector, $n$ is a time-like vector, both normal vectors to the trajectory of the particle $X^{\mu}$, and $S_{\text{constraints}}$ is an action imposing constraints through Langrange multipliers. 
\beq
S_{\text{constraints}}=\int d\sigma \left[ \lambda_1 n\cdot \tilde{n} +\lambda_2 n \cdot \dot{X}+\lambda_3 \tilde{n}\cdot \dot{X} +\lambda_4 \left( n^2+1 \right) +\lambda_5 \left( \tilde{n}^2-1 \right)  \right]\, .
\eeq
The action \eqref{eq:Sspin} introduces two vectors in three-dimensional Minkowski space, amounting to six new degrees of freedom. It also introduces five constraints, resulting in a single new degree of freedom. Studying the equations of motion however reveals that it is not a true degree of freedom. The action \eqref{eq:Sspin} is insensitive to the variation of $n,\tilde{n}$ along the path $X^{\mu}$, and only depends on the boundary values of the vectors. It is also important to mention that the new equation of motion for $X^{\mu}$ reads
\beq\label{eq:EOMX}
\nabla\left[  \xi \dot{X}^{\mu}+\Delta \dot{X}_{\rho}\nabla \left( n^{\mu}\tilde{n}^{\nu}-n^{\nu}\tilde{n}^{\mu}  \right)  \right]=-{{\Delta}\over 2} \dot{X}^{\nu}\left( n^{\rho}\tilde{n}^{\sigma}-n^{\sigma}\tilde{n}^{\rho}  \right)R^{\mu}_{\\ \nu\rho\sigma}\, .
\eeq
Using the constraint equations one can show that  $\tilde{n}$ is completely fixed as a function of $n$. Namely, $\tilde{n}^{\mu}=\epsilon^{\mu \nu \rho}\dot{X}_{\nu}n_{\rho}$. This, together with the fact that in Minkowski space the Riemann tensor vanishes, is enough to conclude that the solution to \eqref{eq:EOMX} is still a straight line obeying $\ddot{X}^{\mu}=0$. This is equivalent to the statement that particles with spin also move in straight lines in a Minkowski background. The on-shell version of the action \eqref{eq:Sspin} of a particle moving from point $x_i$ and vector $n_i$ to a point $x_f$ with vector $n_f$ reads
\beq\label{eq:SspinOS}
S_{\text{on-shell}}=\xi L(x_i,x_f)+\Delta \cosh^{-1}\left(- n_i\cdot n_f \right)\, ,
\eeq
where $L(x_i,x_f)$ is the length of a straight line connecting the points $x_i$ and $x_f$, while $n_i\cdot n_f$ is obtained by parallel transporting $n_f$ from $x_f$ to $x_i$ along $X^{\mu}$ and taking the scalar product between the resulting vector and $n_i$. Otherwise, one can compute $n_i\cdot n_f$ by writing the vectors in global Minkowski coordinates and taking the scalar product directly. The second term in the action corresponds to the total boost parameter required to take the frame defined by $(\dot X, n, \tilde n)$ at the initial point to that at the final point.

In order to compute low-point correlators and Poincar\'e blocks involving spinning particles, we need to repeat the method of previous sections using the new on-shell action \eqref{eq:SspinOS}. As in previous sections, we regulate the problem by translating the boundary points to new bulk points $y_i$ that lie on null lines $\gamma_i$ falling from the boundary. The new ingredient is that we must introduce bulk vectors $n$ at each of the bulk points. A natural assignment would be to use the direction of the null ray  assigned to each boundary point. We thus define
\beq
\dot{\gamma}_i = \partial_r\vert_{\gamma_i}\, .
\eeq
 The extrimization of $r_i$ along $\gamma_i$ ensures that $\dot{\gamma}_i$ is orthogonal to the outgoing line, so that this definition will not result in a degenerate frame. However, the action above only makes sense for $n_i$ timelike. One way to think of this is that the boost required to take us between different null vectors is infinite and must be regulated. We will do so by introducing a regulator $\epsilon$ and taking $n_i$ to approach $\dot{\gamma}_i$ in the limit where $\epsilon \rightarrow 0$, but keeping $n_i$ a finite timelike unit vector for finite $\epsilon$. The vector associated to the regulated bulk point at $\gamma_i$ is then
 \begin{align}\label{eq:mi}
n_i = \frac{1}{\epsilon} \dot{\gamma}_i + \epsilon m_i\, , \quad \text{where $m_i$ is null and }\gamma_i \cdot m_i = -1\, .
\end{align}
There is another perspective from which this prescription is natural. When this action was used in the standard AdS/CFT setup \cite{Castro:2014tta}, the timelike vector defining the frame at the boundary was chosen to coincide with the boundary coordinate time direction, $\partial_t$. Similarly here, we could start with the vector $\partial_u$ at large $r$ and parallel transport it down the null ray to the bulk point $y_i$. This will give us a null vector orthogonal to $\dot \gamma$, which corresponds to the $\epsilon \rightarrow0$ limit of $\tilde n$ in the  prescription \ref{eq:mi}.

Bulk vertices also come equipped with a unit time-like vector $n_{\text{vertex}}$ for which the on-shell action must be extremized.

Note that the dependence of the on-shell action \eqref{eq:Sspin} on the vectors $n$ and the points $x$ factorizes, and so the extremizations with respect to the bulk points and the bulk vectors are completely independent. We can then borrow the results from previous sections for the $\xi$-dependent part of the correlators/blocks, and focus on extremizing the total on-shell action involving the vectors $n$ to obtain the $\Delta$-dependent parts.

We now have all necessary ingredients to compute and extremize the spinning part of the total on-shell action. The total action will be composed of terms that connect the regulated bulk vectors with vertex vectors, and terms that connect vertex vectors with vectors at other vertices. These can be written  respectively as
\beq\label{eq:terms}
\cosh^{-1}\left( -n_i \cdot n_{v} \right)\, , \quad \text{and}\quad \cosh^{-1}\left(-n_{v} \cdot n_{v'} \right)\, .
\eeq
The first object will diverge as $n_i$ becomes null in the $\epsilon\rightarrow 0$ limit of equation \ref{eq:mi}. The terms in the action which connect these regulated bulk vectors to other bulk vectors have the form
\begin{align}
\cosh^{-1}\left( -n_i \cdot n_{v} \right) = \log\frac{-2\dot{\gamma}_i \cdot n_v}{\epsilon} + O(\epsilon) \,,
\end{align}
independent of the precise choice of $m_i$ used to regulate. This contribution to the action is divergent, but can be regulated by adding a counter-term of the form $S_{ct}=\Delta \log \epsilon$ for each boundary point. In what follows we will directly use this regulated action for such segments.

As we will show in the subsections below, extremizing a total on-shell action composed of these terms gives rise to the $\Delta$-dependent part of BMS$_3$ correlators and Poincar\'e blocks in the $1 \ll \Delta \ll \frac{1}{G}$ probe limit.

Before moving on to explicit calculations, it is worth mentioning a particularly useful connection between the total on-shell action at hand and the holographic calculation of correlators and global blocks of CFT$_1$ primaries.

\subsection{Recasting the spinning calculation as holographic blocks in AdS$_2$/CFT$_1$}
Note that the $\Delta$-dependent part of BMS$_3$ correlators are identical to those appearing in a CFT$_1$, since they must both by invariant under the same Virasoro algebra generated by the $\mathcal{L}_n$. 

 We start by choosing an illuminating parametrization of a general time-like vector attached to each of the bulk points
\beq\label{eq:TimeLikeVector}
n=\sec \rho \partial_t +\tan\rho  \cos \phi  \partial_x+\tan\rho \sin \phi\partial_y\, .
\eeq
Null vectors can also be written in this form. In the limit $\rho\rightarrow \pi/2$, the vector in \eqref{eq:TimeLikeVector} becomes null
\beq
m=\lim_{\cos\rho\rightarrow 0}n={1\over{\cos\rho}}\left(
 \partial_t 
+\cos \phi  \partial_x
+\sin \phi \partial_y 
\right)\, .
\eeq
This null vector matches the $\epsilon\rightarrow 0$ limit of \eqref{eq:mi} if we identify $\cos \rho$ with $\epsilon$, as well as $\phi$ with $\phi^{\partial}_i$.  

The key point is that vectors of the form \eqref{eq:TimeLikeVector} can be thought as points in the 2+1 dimensional embedding space of global euclidean AdS$_2$ \cite{Costa:2014kfa}. Because they are time-like, they correspond to bulk points inside AdS$_2$. The null vectors $m$ correspond to points at the boundary. Note that $\cosh^{-1}\left( -n \cdot n' \right)$ corresponds to the geodesic distance between the AdS$_2$ points associated with $n$ and $n'$. We conclude that the total spinning on-shell action we need to extremize corresponds to the length of a network of geodesic segments in euclidean global AdS$_2$.

 In \cite{Hijano:2015rla} it was shown that such objects compute holographic correlators and global blocks of primaries in the dual conformal field theory, in the limit of large central charge $c\rightarrow \infty$ with scaling dimension $\Delta/c=\epsilon\ll 1$. This limit is also equivalent to first taking the large central charge limit while keeping the scaling dimensions $\Delta\sim{\cal O}(1)$, and then sending the scaling dimensions to $\infty$ \cite{Hijano:2015zsa}. We can then borrow euclidean AdS$_2$ results to obtain the $\Delta$-dependent part of BMS$_3$ correlators and global blocks.

The results for low-point correlators and conformal blocks in CFT$_1$ can be found in \cite{Chamon:2011xk,Jackiw:2012ur} in coordinates associated to the boundary of Lorentzian AdS$_2$ in Poincar\'e coordinates
\beq\label{eq:Ppatch}
ds^2={{dz^2-dt^2}\over{z^2}}\, ,
\eeq
where $z$ is the holographic coordinate and the CFT$_1$ is located at the $z=0$ boundary. The geodesic networks obtained from our spinning on-shell action correspond to euclidean global AdS$_2$ with line element
\beq
ds^2={{d\rho^2+\sin^2\rho d\phi^2}\over{\cos^2\rho}}\, ,
\eeq
where the boundary is now located at $\rho\rightarrow \pi/2$.  The topology is that of a disk, whose standard map to the euclidean infinte strip reads
\beq\label{eq:CFT1map}
\sigma +i\tau = 2\tan^{-1} \tanh{{\log\tan{{\rho}\over 2}+i\phi}\over 2}\, . 
\eeq
The line element in the new coordinates reads
\beq
ds^2 = {{d\sigma^2+d\tau^2}\over {\sin^2\sigma}}\, . 
\eeq
Under this map, the boundary of the disk with $-\pi/2<\phi<\pi/2$ is mapped to the line of constant $\sigma=0$, while the boundary of the disk with $\pi/2<\phi<3\pi/2$ is mapped to the line of constant $\sigma=\pi$. Euclidean global AdS$_2$ is thus equivalent to two copies of a CFT$_1$ on a line. 

The usual state-operator correspondence on a $d$-dimensional conformal field theory relates the action of local operators in $S^d$ to states in a conformal field theory on $S^{d-1}\times \mathbb{R}$. In the particular case of $d=1$, the map relates the action of local operators in $S^1$ (boundary of a disk), to states in $S^0 \times \mathbb{R}$. The manifold $S^0$ is simply two points, which in our choice of coordinates correspond to the $\sigma=0$ and the $\sigma=\pi$ points. Equation \eqref{eq:CFT1map} at the boundary is nothing but the explicit map between $S^1$ and $S^0\times \mathbb{R}$.

\begin{figure}[t!]
\centering

\begin{tikzpicture}
\draw[black, very thick,fill={rgb:black,1;white,4}] (2,2) circle (2)  pic[black, very thick, -stealth]{carc=0:30:1.8} pic[red, very thick]{carc=90:270:2}  pic[blue, very thick]{carc=-90:90:2};

\draw[black, very thick,-stealth] (2,2)--({2+2/sqrt(2)},{2+2/sqrt(2)}) node [pos=0.5,left] {$\rho$} ;

\draw (3.5,2.6) node[below] {$\phi$};

\draw[black, very thick,-latex] (4.5,2)--(5.5,2) node[pos=0.5,above]{(a)};

\fill [fill={rgb:black,1;white,4}] (6,0) rectangle (8,4);
\draw[red,very thick] (6,0)--(6,4) node [pos=0,below] {\textcolor{black}{$\sigma=0$}};
\draw[blue,very thick] (8,0)--(8,4) node [pos=0,below=2] {\textcolor{black}{$\sigma=\pi$}};

\draw[black, very thick,-stealth] (7.7,1.5)--(7.7,2.5) node [pos=0.5,left] {$\tau$} ;
\draw[black, very thick,-stealth] (6.5,3.6)--(7.5,3.6) node [pos=0.5,above] {$\sigma$} ;

\draw[black, very thick,-latex] (8.5,2)--(9.5,2)  node[pos=0.5,above]{(b)};

\fill [fill={rgb:black,1;white,4}] (10,0) rectangle (12,4);
\fill[green, opacity=0.2] (12,1)--(10,2)--(12,3)--cycle;

\draw[black, dashed] (10,0)--(12,1);
\draw[black, dashed] (12,1)--(10,2);
\draw[black, dashed] (10,2)--(12,3);
\draw[black, dashed] (12,3)--(10,4);

\draw[red,very thick] (10,0)--(10,4) node [pos=0,below] {\textcolor{black}{$\sigma=0$}};
\draw[blue,very thick] (12,0)--(12,4) node [pos=0,below=2] {\textcolor{black}{$\sigma=\pi$}};

\draw[black, very thick,-stealth] (11.7,1.5)--(11.7,2.5) node [pos=0.5,left] {$T$} ;
\draw[black, very thick,-stealth] (10.5,3.6)--(11.5,3.6) node [pos=0.5,above] {$\sigma$} ;

\draw[black, very thick,-latex] (12.5,2)--(13.5,2)  node[pos=0.5,above]{(c)};

\draw[blue,very thick] (14,0)--(14,4) node [pos=0,below=2] {\textcolor{black}{$z=0$}};

\fill[green,left color=green!20,right color=white!10] (14,0)--(14,4)--(16,4)--(16,0)--cycle;

\draw[black, very thick,-stealth] (14.3,1.5)--(14.3,2.5) node [pos=0.5,right] {$t$} ;
\draw[black, very thick,-stealth] (14,3.6)--(15,3.6) node [pos=0.5,above] {$z$} ;

\end{tikzpicture}

    \caption{Map from euclidean global AdS$_2$ (left) to the Poincar\'e patch (right). a) Transformation from the disk to the strip, implemented by the change of coordinates in equation \eqref{eq:CFT1map}. b) Inverse Wick rotation from the euclidean strip to the lorentzian one, implemented by $\tau=i T$. c) Poincar\'e patching of the lorentzian strip, implemented by formula \eqref{eq:PoincareMap}. }\label{fig:MAP}
\end{figure}
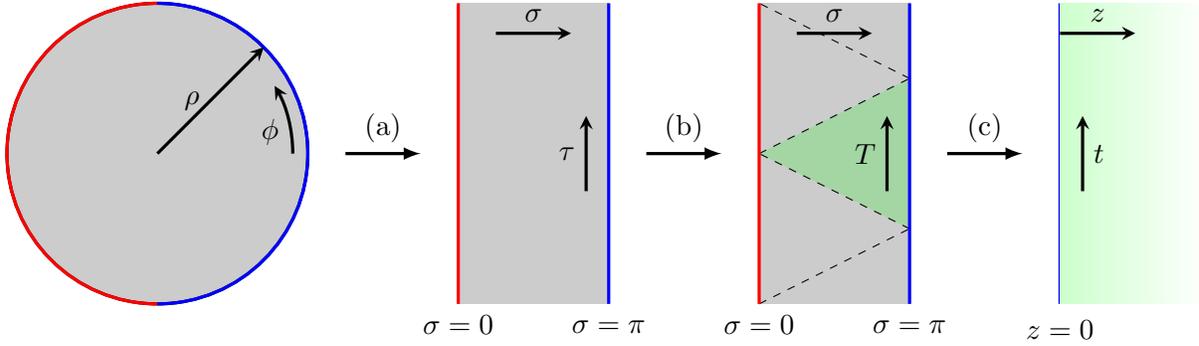

The Euclidean strip can be inverse-Wick rotated to Lorentzian signature by sending $\tau = i T$. The lorentzian strip is then covered with Poincar\'e patches with coordinates
\beq\label{eq:PoincareMap}
z={{\sin\sigma}\over{\cos T+\cos\sigma}}\, , \quad t={{\sin T}\over{\cos T+\cos \sigma}}\, ,
\eeq
which give rise to the line element in equation \eqref{eq:Ppatch}. All in all, the map between  Poincar\'e time and the angular disk coordinate  at the boundary reads
\beq\label{eq:Bmap}
\begin{split}
t&=-i \tan{{\phi}\over 2}\, , \quad \text{for} \quad \phi\in \left(-{{\pi}\over 2},{{\pi}\over 2}\right)\, .\\
t&=-i \cot{{\phi}\over 2}\, , \quad \text{for} \quad \phi\in \left({{\pi}\over 2},{{3\pi}\over 2}\right)\, .
\end{split}
\eeq
For a picture that  summarizes this construction, see figure \ref{fig:MAP}. In the following subsections, we will make use of this map to compute the spinning part of correlators and blocks in a BMSFT.

 With this guarantee that our prescription produces the desired results, we turn now to explicit computations of spinning two- and three-point correlators and Poincar\'e blocks. 

\subsection{Spinning two-point function}
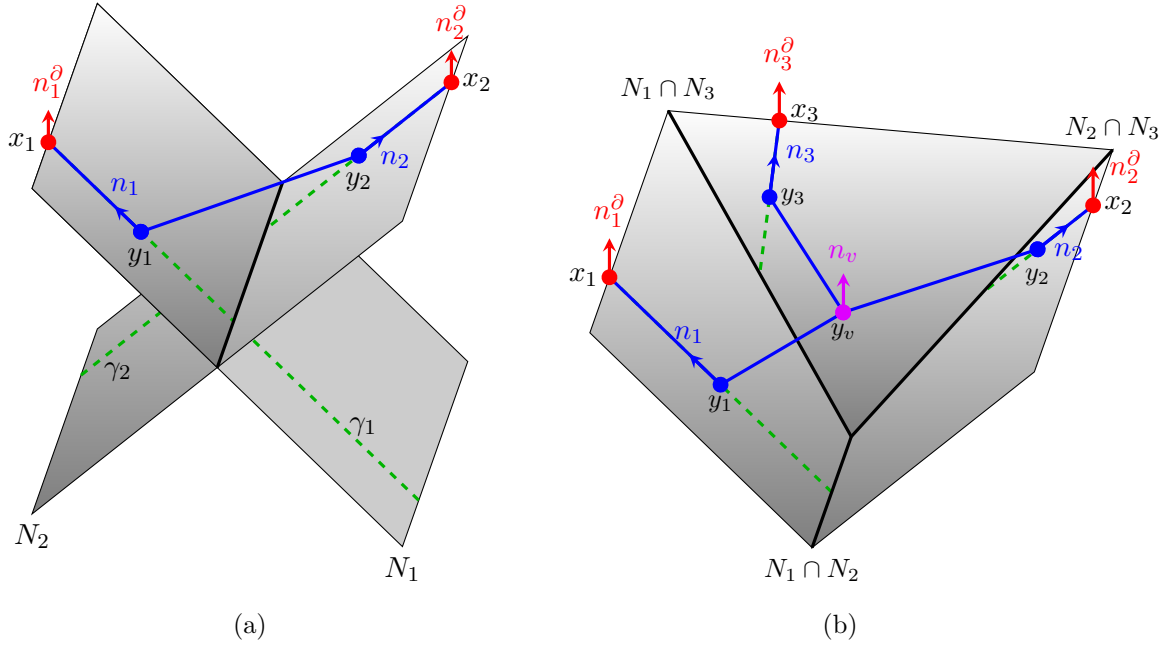
\begin{figure}[t!]
\centering
\begin{subfigure}[t]{0.48\textwidth}
        \centering
\tdplotsetmaincoords{60}{100}
\begin{tikzpicture}[scale=5,tdplot_main_coords,vectorR/.style={-stealth,red,very thick},vectorB/.style={-stealth,blue,very thick}]

    \def\x{1}

  \filldraw[
        draw=black,%
        fill=black!20,%
    ]          (0,1,0)
            -- (1,1,0)
            -- (1,1/2,1/2)
            -- (0,1/2,1/2)
            -- cycle;

  \draw[
        darkgreen,dashed,very thick%
    ]          (3/4,1,0)
            -- (3/4,1/2,1/2) node [pos=0.3,above,name=int] {{\textcolor{black}{$\gamma_1$}}};

    \filldraw[
        draw=black,%
        fill=black!20, bottom color=gray, top color=white%
    ]          (0,0,0)
            -- (1,0,0)
            -- (1,1,1)
            -- (0,1,1)
            -- cycle;

  \draw[
        darkgreen,dashed,very thick%
    ]          (1/4,0,0)
            -- (1/4,1/2,1/2) node [pos=0.2,below=2,name=int] {{\textcolor{black}{$\gamma_2$}}};

  \filldraw[
        draw=black,%
        fill=black!20,bottom color=gray, top color=white%
    ]          (0,0,1)
            -- (1,0,1)
            -- (1,1/2,1/2)
            -- (0,1/2,1/2)
            -- cycle;

  \filldraw[
        draw=black,%
        fill=black!20,bottom color=gray, top color=white%
    ]          (0,0,1)
            -- (1,0,1)
            -- (1,1/2,1/2)
            -- (0,1/2,1/2)
            -- cycle;

\draw (1,0,0) node[below] {$N_2$};
\draw (1,1,0) node[below] {$N_1$};

  \draw[
         darkgreen,dashed,very thick%
    ]          (3/4,0,1)
            -- (3/4,1/2,1/2) node [pos=0,left,name=O1] {\textcolor{black}{$x_1$}};
  \draw[
          darkgreen,dashed,very thick%
    ]          (1/4,1,1)
            -- (1/4,1/2,1/2) node [pos=0,right,name=O2] {\textcolor{black}{$x_2$}};

  \draw[
       black,very thick%
    ]          (0,1/2,1/2)
            -- (1,1/2,1/2);

  \draw[
         -dot-=0,blue,very thick%
    ]          (3/4,0,1)
            -- (3/4,1/4,3/4);

  \draw[
        -dot-=0, blue,very thick%
    ]          (1/4,1,1)
            -- (1/4,3/4,3/4);

  \draw[
       -dot2-=0, -dot2-=1,blue,very thick%
    ]         (3/4,1/4,3/4)
            -- (1/4,3/4,3/4)   node [pos=1,below=2,name=OB2] {\textcolor{black}{$y_2$}}  node [pos=0,below=2,name=OB1] {\textcolor{black}{$y_1$}};
            
\draw[vectorR]   (3/4,0,1) --   (3/4+0,0,1+0.1) node [pos=1,left,above] {$n^{\partial}_1$};
\draw[vectorR]  (1/4,1,1)--   (1/4+0,1,1+0.1)  node [pos=1,right,above] {$n^{\partial}_2$};

\draw[vectorB]  (3/4,1/4,3/4) --   (3/4,{1/4-0.1/sqrt(2)},{3/4+0.1/sqrt(2)}) node [pos=1,right=4,above] {$n_1$};
\draw[vectorB]  (1/4,3/4,3/4) --   (1/4,{3/4+0.1/sqrt(2)},{3/4+0.1/sqrt(2)}) node [pos=1,right=4,below=2] {$n_2$};

\end{tikzpicture}
        \caption{ }
    \end{subfigure}%
\quad
\begin{subfigure}[t]{0.48\textwidth}
        \centering

\tdplotsetmaincoords{60}{100}
\begin{tikzpicture}[scale=6,tdplot_main_coords,vectorR/.style={-stealth,red,very thick},vectorB/.style={-stealth,blue,very thick},vectorP/.style={-stealth,purple2,very thick}]

    \def\x{1}

    \filldraw[
        draw=black,%
        fill=black!20,bottom color=gray, top color=white%
    ]          (0,1/2,0)
            -- (1,1/2,0)
            -- (1,1,1/2)
            -- (0,1,1/2)
            -- cycle;

  \filldraw[
        draw=black,%
        fill=black!20,bottom color=gray, top color=white%
    ]          (0,0,1/2)
            -- (1,0,1/2)
            -- (1,1/2,0)
            -- (0,1/2,0)
            -- cycle;

  \draw[
         darkgreen,dashed,very thick%
    ]          (3/4,0,1/2)
            -- (3/4,1/2,0) node [pos=0,left,name=O1] {\textcolor{black}{$x_1$}};
  \draw[
          darkgreen,dashed,very thick%
    ]          (1/4,1,1/2)
            -- (1/4,1/2,0) node [pos=0,right,name=O2] {\textcolor{black}{$x_2$}};

  \draw[
       black,very thick%
    ]          (0,1/2,0)
            -- (1,1/2,0)  node [pos=1,below,name=int] {{\small\textcolor{black}{$N_1\cap N_2$}}};

  \filldraw[
        draw=black,%
        fill=black!20,bottom color=gray, top color=white%
    ]          (1/2,1/2,0)
            -- (0,1,1/2)
            -- (0,0,1/2)
            -- cycle;

  \draw[
          darkgreen,dashed,very thick%
    ]          (0,1/4,1/2)
            -- (1/4,1/4,1/4) node [pos=0,above=3,name=O3, right] {\textcolor{black}{$x_3$}};

  \draw[
       black,very thick%
    ]          (1/2,1/2,0)
            -- (0,1,1/2)  node [pos=1,above,name=int] {{\small\textcolor{black}{$N_2\cap N_3$}}};

  \draw[
       black,very thick%
    ]          (1/2,1/2,0)
            -- (0,0,1/2)  node [pos=1,above,name=int] {{\small\textcolor{black}{$N_1\cap N_3$}}};

  \draw[
         -dot-=0,blue,very thick%
    ]          (3/4,0,1/2)
            -- (3/4,1/4,1/4);

  \draw[
         -dot-=0,blue,very thick%
    ]          (0,1/4,1/2)
            -- (1/8,1/4,{1/2-1/8});

  \draw[
         -dot-=0,blue,very thick%
    ]          (1/4,1,1/2)
            -- (1/4,7/8,{1/2-1/8});

  \draw[
         -dot2-=0,blue,very thick%
    ]          (3/4,1/4,1/4)
            -- (1/2+0.1,1/2,3/8)  node [pos=0,below,name=int] {{\small\textcolor{black}{$y_1$}}};

  \draw[
         -dot2-=0,blue,very thick%
    ]         (1/8,1/4,{1/2-1/8})
            --(1/2+0.1,1/2,3/8)   node [pos=0,right,name=int] {{\small\textcolor{black}{$y_3$}}};

  \draw[       -dot2-=0, -dot3-=1,blue,very thick ]         (1/4,7/8,{1/2-1/8}) -- (1/2+0.1,1/2,3/8) node [pos=1,below,name=int] {{\small\textcolor{black}{$y_v$}}}  node [pos=0,below=3,name=int] {{\small\textcolor{black}{$y_2$}}};

\draw[vectorR]   (3/4,0,1/2) --   (3/4+0,0,1/2+0.1) node [pos=1,left,above] {$n^{\partial}_1$};
\draw[vectorR]   (0,1/4,1/2) --   (0+0,1/4,1/2+0.1) node [pos=1,left,above] {$n^{\partial}_3$};
\draw[vectorR]   (1/4,1,1/2) --   (1/4+0,1,1/2+0.1) node [pos=1,left,right=2] {$n^{\partial}_2$};

\draw[vectorB]     (3/4,1/4,1/4) --    (3/4,{1/4-0.1/sqrt(2)},{1/4+0.1/sqrt(2)})node [pos=1,above] {$n_1$};
\draw[vectorB]     (1/8,1/4,{1/2-1/8}) --    ({1/8-0.1/sqrt(2)},1/4,{1/2-1/8+0.1/sqrt(2)}) node [pos=1,right] {$n_3$};
\draw[vectorB]     (1/4,7/8,{1/2-1/8}) --    (1/4,{7/8+0.1/sqrt(2)},{1/2-1/8+0.1/sqrt(2)}) node [pos=1,below=3] {$n_2$};

\draw[vectorP]    (1/2+0.1,1/2,3/8) --   ({1/2+0.1+0/sqrt(3)},{1/2+0/sqrt(3)},{3/8+0.1}) node [pos=1,above] {$n_{v}$};

\end{tikzpicture}
    
        \caption{ }
    \end{subfigure}
    \caption{a) Set-up for the calculation of the holographic spinning two-point function. The red points are located at the null boundary at $x_{i=1,2}$, equipped with  boundary vectors $n^{\partial}_{i}=\partial_u$. The regulated blue points are located at $\gamma_{i}$, and they are equipped with blue vectors $n_{i}=\partial_r\vert_{y_{i}}$. The spinning particle propagates between the blue points, along the green straight line connecting $\gamma_1$ and $\gamma_2$.  b) Set-up for the calculation of the holographic spinning three-point function. There are three boundary points drawn in red, and their regulated bulk versions are drawn in blue. At each blue point we have a blue vector $n_{i=1,2,3}=\partial_r\vert_{y_{i}}$ pointing along the null geodesics $\gamma_i$. The spining particles meet at a common purple point $y_v$, where we have a purple time-like vector $n_v$ for which the spinning total on-shell action must be extremized.}\label{fig:lptSpin}
\end{figure}

A figure of the set-up involved in this section can be found in Figure \ref{fig:lptSpin}.a). In the case of a two-point function we have two bulk points $y_1$ and  $y_2$ at the locations $N_1\cap N_2\cap \gamma_1$ and $N_1\cap N_2\cap \gamma_2$ respectively. At each of these points we have a regulated bulk vector pointing in the radial direction (along the null lines $\gamma_i$)
\beq\label{eq:n12}
\dot{\gamma}_1=\partial_{r}\vert_{y_1}=\partial_t+\cos\phi^{\partial}_1\partial_x+\sin\phi^{\partial}_1\partial_y\, , \quad  \dot{\gamma}_2=\partial_{r}\vert_{y_2}=\partial_t+\cos\phi^{\partial}_2\partial_x+\sin\phi^{\partial}_2\partial_y\,  .
\eeq
These two vectors are automatically perpendicular to the path connecting $y_1$ and  $y_2$, which lies on $N_1\cap N_2$. However, these vectors are null, while the action \eqref{eq:Sspin} concerns time-like vectors. For this reason we introduce the following time-like vectors, as explained above equation \eqref{eq:mi}
\beq
n_1 ={1\over {\epsilon}} \dot{\gamma}_1-{{\epsilon}\over{2}} {{1}\over{\dot{\gamma}_1 \cdot \dot{\gamma}_2}} \dot{\gamma}_2\, , \quad n_2={1\over {\epsilon}} \dot{\gamma}_2- {{\epsilon}\over{2}}{{1}\over{\dot{\gamma}_1 \cdot \dot{\gamma}_2}} \dot{\gamma}_1\, ,
\eeq
which become null and proportional to \eqref{eq:n12} in the $\epsilon\rightarrow 0$ limit. The spinning part of the on-shell action is then
\beq
S^{\text{spin, regulated}}_{\text{total}}=\Delta \lim_{\epsilon\rightarrow 0} \left[\cosh^{-1}\left(-n_1\cdot n_2 \right)+2\Delta\log\epsilon \right]= 2\Delta \log \left({\dot{\gamma}_1 \cdot \dot{\gamma}_2}\right) = 2\Delta \log {{2\sin{{\phi^{\partial}_2-\phi^{\partial}_1}\over 2}}}\, .
\eeq
The contribution to the correlator is then 
\beq
e^{-S^{\text{spin, reg.}}_{\text{total}}}=\left({{2 \sin {{\phi^{\partial}_2-\phi^{\partial}_1}\over 2}}}\right)^{-2\Delta}\, ,
\eeq
which combined with the $\xi$-dependent part computed in section \ref{sec:2pt} gives rise to the correct result as written in equation \eqref{eq:lowpt}. 

As we have explained in previous subsections, we could have obtained this same answer by borrowing results from computations in a CFT$_1$. The two point function of primary operators reads
\beq
\langle   \Phi_{\Delta}(t_1)  \Phi_{\Delta}(t_2)   \rangle\sim {1\over{(t_2-t_1)^{2\Delta}}}\, .
\eeq
Performing the change of coordinates of equation \eqref{eq:Bmap} yields
\beq
\langle   \Phi_{\Delta}(\phi^{\partial}_1)  \Phi_{\Delta}(\phi^{\partial}_2)   \rangle\sim \left( {{\partial\tan{{\phi_1}\over 2}}\over{\partial \phi_1}} \right)^{\Delta}\left( {{\partial\tan{{\phi_2}\over 2}}\over{\partial \phi_2}} \right)^{\Delta} {1\over{(\tan{{\phi_2}\over 2}  -\tan{{\phi_1}\over 2})^{2\Delta}}}=\left(  {1\over {2\sin {{\phi^{\partial}_{21}}\over 2}}} \right)^{2\Delta}\, ,
\eeq
which matches the result we have obtained by extremizing the spinning on-shell action.

\subsection{Spinning three-point function}
The geometrical set-up studied in this section is presented in figure \ref{fig:lptSpin}.b). We start with three regulated bulk points $y_i$ along the $\gamma_i$ associated to each boundary point connected to a bulk vertex at $y_v$. The action to be extremised is: 
\beq
S^{\text{spin, regulated}}_{\text{total}}= \sum^3_{i=1} \Delta_i \log \left(-2 \dot \gamma_i \cdot n_v \right) \, .
\eeq

The problem can be solved in two ways. First, we will solve the extremization problem explicitly. Then, we will obtain the same result by borrowing the result for the analytically continued  euclidean CFT$_1$ three-point function of primaries.

A useful way of parametrizing the vector $n_v$ is the following. In global three-dimensional Minkowski coordinates we have
\beq\label{eq:nvSpin}
n_v={{x^2+u^2+1}\over{2u}}\partial_t +{{x^2+u^2-1}\over{2u}}\partial_x+{x\over u}\partial_y\, ,
\eeq
where $x$ and $u$ are free parameters and the form \eqref{eq:nvSpin} ensures $n_v\cdot n_v=-1$. The interpretation of $x,u$ is simply the Poincar\'e coordinates of euclidean AdS$_2$. The action now reads
\beq
S^{\text{spin}}_{\text{total}}=\sum^3_{i=1} \left[ \Delta_i \log{1\over{\epsilon}} +\Delta_i \log \left( {{u^2+x^2+1}\over{u}}  -{{u^2+x^2-1}\over{u}}  \cos\phi^{\partial}_i-{{2x}\over{u}}  \sin \phi^{\partial}_i\right)   \right]\, .
\eeq
We proceed by changing variables as
\beq
 {{u^2+x^2+1}\over{u}}  -{{u^2+x^2-1}\over{u}}  \cos\phi^{\partial}_i-{{2x}\over{u}}  \sin \phi^{\partial}_i=l_i\, , \quad\text{for} \quad  i=1,2,3\, ,
\eeq
which can be thought as a system of equations for the variables $u$, $x$, and $l_3$. The on-shell action as a function of $l_1$ and $l_2$ reads
\beq
\begin{split}
S^{\text{spin}}_{\text{total}}&=\sum^3_{i=1}  \Delta_i \log{1\over{\epsilon}} +\Delta_1 \log l_1 +\Delta_2 \log l_2\\
&+\Delta_3 \log\left( 
l_1 {{\sin^2 {{\phi^{\partial}_{32}}\over 2}}\over{\sin^2{{\phi^{\partial}_{21}}\over 2}}} 
+l_2 {{\sin^2 {{\phi^{\partial}_{31}}\over 2}}\over{\sin^2{{\phi^{\partial}_{21}}\over 2}}} 
-4\sin{{\phi^{\partial}_{31}}\over 2} \sin{{\phi^{\partial}_{32}}\over 2}\sqrt{l_1 l_2-4\sin^2{{\phi^{\partial}_{21}}\over 2}}
\right)\, .
\end{split}
\eeq
This can now be organized further by replacing
\beq
l_1 = c_1 {{\sin{{\phi^{\partial}_{21}}\over 2} \sin{{\phi^{\partial}_{31}}\over 2}}\over{\sin{{\phi^{\partial}_{32}}\over 2}}}\, , \quad l_2= c_2 {{\sin{{\phi^{\partial}_{21}}\over 2} \sin{{\phi^{\partial}_{32}}\over 2}}\over{\sin{{\phi^{\partial}_{31}}\over 2}}}\, ,
\eeq
which implies
\beq
\begin{split}
S^{\text{spin}}_{\text{total}}&=\sum^3_{i=1}  \Delta_i \log{1\over{\epsilon}} +\Delta_1 \log {{\sin{{\phi^{\partial}_{21}}\over 2} \sin{{\phi^{\partial}_{31}}\over 2}}\over{\sin{{\phi^{\partial}_{32}}\over 2}}}  +\Delta_2 \log {{\sin{{\phi^{\partial}_{21}}\over 2} \sin{{\phi^{\partial}_{32}}\over 2}}\over{\sin{{\phi^{\partial}_{31}}\over 2}}} +\Delta_3 \log {{\sin{{\phi^{\partial}_{31}}\over 2} \sin{{\phi^{\partial}_{32}}\over 2}}\over{\sin{{\phi^{\partial}_{21}}\over 2}}}\\
&+ \Delta_1\log c_1+\Delta_2 \log c_2+\Delta_3 \log \left(  c_1+c_2-2\sqrt{c_1c_2-4} \right)\, .
\end{split}
\eeq
Extremizing now implies solving for $\partial S^{\text{spin}}_{\text{total}}/\partial c_1=\partial S^{\text{spin}}_{\text{total}}/\partial c_2=0$. The solution only contributes to an overall constant in the on-shell action, and it reads
\beq
{{c_1}\over{\sqrt{c_1c_2-4}}}=2{{\Delta_1}\over{\Delta_{132}}}\, , \quad {{c_2}\over{\sqrt{c_1c_2-4}}}=2{{\Delta_2}\over{\Delta_{231}}}\,  .
\eeq
The spinning part of the three-point function thus reads
\beq\label{eq:result3ptspin}
e^{-S^{\text{spin}}_{\text{total}}}=C \prod_k\left( \sin {{\phi^{\partial}_i-\phi^{\partial}_j}\over 2} \right)^{-\Delta_{ijk}} \, ,
\eeq
where $C$ is a constant that depends on $\epsilon$ and $\Delta_i$. As can be seen easily, this matches the $\Delta$-dependent part of the BMS$_3$ three-point function written in \eqref{eq:lowpt}. 

A much quicker way to solve the problem is to  study the CFT$_1$ three-point function of primary operators. On the line, these read
\beq
\langle  \Phi_{\Delta_1}(t_1)  \Phi_{\Delta_2}(t_2)   \Phi_{\Delta_3}(t_3)  \rangle_{\text{CFT}_1} \sim \prod_k \left( t_{ij}\right)^{-\Delta_{ijk}}\, .
\eeq
Performing the change of coordinates from Poincar\'e time to $S^1$ of equation \eqref{eq:Bmap} results in 
\beq
\langle  \Phi_{\Delta_1}(\phi^{\partial}_1)  \Phi_{\Delta_2} (\phi^{\partial}_2)  \Phi_{\Delta_3}(\phi^{\partial}_3)  \rangle_{\text{CFT}_1} \sim \prod_k\left( \sin {{\phi^{\partial}_i-\phi^{\partial}_j}\over 2} \right)^{-\Delta_{ijk}}\, .
\eeq
which of course agrees with the previously obtained result \eqref{eq:result3ptspin}.

\subsection{Spinning Poincar\'e blocks}
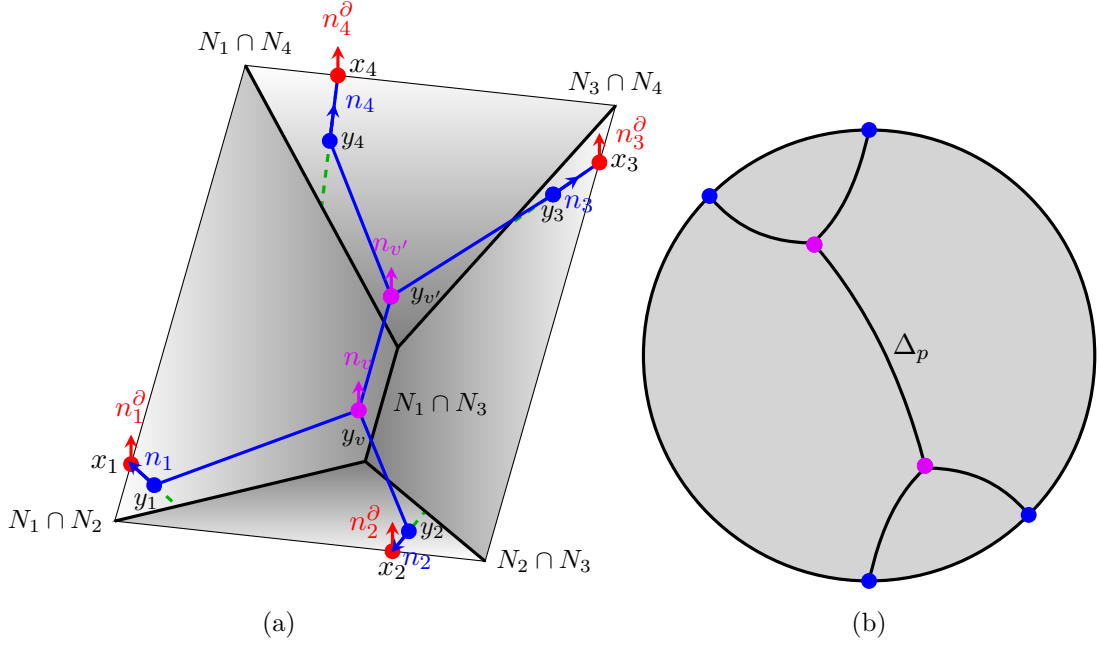
\begin{figure}[t!]
\centering
\begin{subfigure}[t]{0.48\textwidth}
        \centering

\tdplotsetmaincoords{52}{100}
\begin{tikzpicture}[scale=5,tdplot_main_coords,vectorR/.style={-stealth,red,very thick},vectorB/.style={-stealth,blue,very thick},vectorP/.style={-stealth,purple2,very thick}]

    \def\x{1}

    \filldraw[
        draw=black,%
        fill=black!20,left color=gray,right color=white%
    ]          (0,1/2,0)
            -- (1,1/2,0)
            -- (2,1,1/2)
            -- (0,1,1/2)
            -- cycle;

  \filldraw[
        draw=black,%
        fill=black!20,right color=gray,left color=white%
    ]          (0,0,1/2)
            -- (2,0,1/2)
            -- (1,1/2,0)
            -- (0,1/2,0)
            -- cycle;

  \draw[
         darkgreen,dashed,very thick%
    ]          (1+3/4,0,1/2)
            -- (1+3/4,1/8,1/2-1/8) node [pos=0,left,name=O1] {\textcolor{black}{$x_1$}};
  \draw[
          darkgreen,dashed,very thick%
    ]          (1/4,1,1/2)
            -- (1/4,1/2,0) node [pos=0,right,name=O2] {\textcolor{black}{$x_3$}};

  \draw[
       black,very thick%
    ]          (0,1/2,0)
            -- (1,1/2,0)  node [pos=0.75,right,name=int] {{\small\textcolor{black}{$N_1\cap N_3$}}};

  \filldraw[
        draw=black,%
        fill=black!20,bottom color=gray, top color=white%
    ]          (1/2,1/2,0)
            -- (0,1,1/2)
            -- (0,0,1/2)
            -- cycle;

  \filldraw[
        draw=black,%
        fill=black!20,bottom color=white, top color=gray%
    ]          (1,1/2,0)
            -- (2,0,1/2)
            -- (2,1,1/2)
            -- cycle;

  \draw[
          darkgreen,dashed,very thick%
    ]          (0,1/4,1/2)
            -- (1/4,1/4,1/4) node [pos=0,above=3, right,name=O3] {\textcolor{black}{$x_4$}};

  \draw[
          darkgreen,dashed,very thick%
    ]          (2,3/4,1/2)
            -- ({2-3/6},3/4,{1/2-(3/6)*1/2}) node [pos=0,below,name=O3] {\textcolor{black}{$x_2$}};

  \draw[
       black,very thick%
    ]          (1/2,1/2,0)
            -- (0,1,1/2)  node [pos=1,above,name=int] {{\small\textcolor{black}{$N_3\cap N_4$}}};

  \draw[
       black,very thick%
    ]          (1/2,1/2,0)
            -- (0,0,1/2)  node [pos=1,above,name=int] {{\small\textcolor{black}{$N_1\cap N_4$}}};

  \draw[
       black,very thick%
    ]          (1,1/2,0)
            -- (2,0,1/2)  node [pos=1,left,name=int] {{\small\textcolor{black}{$N_1\cap N_2$}}};

  \draw[
       black,very thick%
    ]          (1,1/2,0)
            -- (2,1,1/2)  node [pos=1,right,name=int] {{\small\textcolor{black}{$N_2\cap N_3$}}};

  \draw[
         -dot-=0,blue,very thick%
    ]          (2-1/4,0,1/2)
            --  (2-1/4,1/16,1/2-1/16);

  \draw[
         -dot-=0,blue,very thick%
    ]          (2,3/4,1/2)
            -- (2-2/8,3/4,1/2-2/16);

  \draw[
         -dot-=0,blue,very thick%
    ]          (0,1/4,1/2)
            -- (0+1/8,1/4,1/2-1/8);

  \draw[
         -dot-=0,blue,very thick%
    ]          (1/4,1,1/2)
            -- (1/4,1-1/8,1/2-1/8);

  \draw[
         -dot2-=0,blue,very thick%
    ]         (2-1/4,1/16,1/2-1/16)
            -- (1/2+1/2+0.1,1/2,1/4)   node [pos=0,left=3,below=-1,name=int] {{\small\textcolor{black}{$y_1$}}};

  \draw[
         -dot2-=0,blue,very thick%
    ]          (2-2/8,3/4,1/2-2/16)
            -- (1/2+1/2+0.1,1/2,1/4)   node [pos=0,right,name=int] {{\small\textcolor{black}{$y_2$}}};

  \draw[
         -dot2-=0,blue,very thick%
    ]        (0+1/8,1/4,1/2-1/8)
            -- (1/2+0.1,1/2,1/4)   node [pos=0,right,name=int] {{\small\textcolor{black}{$y_4$}}};

  \draw[
         -dot2-=0, -dot3-=1,blue,very thick%
    ]         (1/4,1-1/8,1/2-1/8)
            -- (1/2+0.1,1/2,1/4) node [pos=0,below,name=int] {{\small\textcolor{black}{$y_3$}}};

  \draw[
         -dot3-=0, -dot3-=1,blue,very thick%
    ]        (1/2+0.1,1/2,1/4)
            -- (1/2+1/2+0.1,1/2,1/4) node [pos=1,below=3,name=int] {{\small\textcolor{black}{$y_{v}\,\,$}}}  node [pos=0,right=3,name=int] {{\small\textcolor{black}{$y_{v'}$}}};

\draw[vectorR]   (1+3/4,0,1/2) --   (1+3/4+0,0,1/2+0.1) node [pos=1,left,above] {$n^{\partial}_1$};
\draw[vectorR]   (0,1/4,1/2) --   (0+0,1/4,1/2+0.1) node [pos=1,left,above] {$n^{\partial}_4$};
\draw[vectorR]   (1/4,1,1/2) --   (1/4+0,1,1/2+0.1) node [pos=1,left,right=2] {$n^{\partial}_3$};
\draw[vectorR]   (2,3/4,1/2) --   (2,3/4,1/2+0.1) node [pos=1,left] {$n^{\partial}_2$};

\draw[vectorB]     (2-1/4,1/16,1/2-1/16) --    (2-1/4,{1/16-0.1/sqrt(2)},{1/2-1/16+0.1/sqrt(2)})node [pos=1,above,right=2] {$n_1$};
\draw[vectorB]     (0+1/8,1/4,1/2-1/8) --    ({1/8-0.1/sqrt(2)},1/4,{1/2-1/8+0.1/sqrt(2)}) node [pos=1,right] {$n_4$};
\draw[vectorB]     (1/4,1-1/8,1/2-1/8) --    (1/4,{1-1/8+0.1/sqrt(2)},{1/2-1/8+0.1/sqrt(2)}) node [pos=1,below=3] {$n_3$};
\draw[vectorB]        (2-2/8,3/4,1/2-2/16) --   ({2-2/8+0.6/sqrt(5)},3/4,{1/2-2/16+0.3/sqrt(5)}) node [pos=1,below=3,right] {$n_2$};

\draw[vectorP]    (1/2+0.1,1/2,1/4) --   ({1/2+0.1},{1/2+0/sqrt(3)},{1/4+0.1}) node [pos=1,above] { $n_{v'}$};
\draw[vectorP]    (1/2+1/2+0.1,1/2,1/4) --   ({1/2+1/2+0.1+0/sqrt(3)},{1/2+0/sqrt(3)},{1/4+0.1}) node [pos=1,above] {$n_{v}$};

\end{tikzpicture}
    
        \caption{ }
    \end{subfigure}%
\quad
\begin{subfigure}[t]{0.48\textwidth}
        \centering
\begin{tikzpicture}[scale=1.5]
\draw[black, very thick,fill={rgb:black,1;white,5}] (2,2) circle (2);

\draw [black,very thick] plot [smooth, tension=1] coordinates {(2,0)  (2.2,0.6)  (2.5,1)};

\draw [-dot2-=0,black,very thick] (2,0)--(2,0.01);

\draw [black,very thick] plot [smooth, tension=1] coordinates { ({2+2/sqrt(2)},{2-2/sqrt(2)})  ({2+2/sqrt(2)-0.4},0.9)  (2.5,1)};
\draw [-dot2-=0,black,very thick] ({2+2/sqrt(2)},{2-2/sqrt(2)})-- ({2+2/sqrt(2)},{2-2/sqrt(2)+0.001});

\draw [black,very thick] plot [smooth, tension=1] coordinates { ({2-2/sqrt(2)},{2+2/sqrt(2)}) ({2-2/sqrt(2)+0.4},3.1)   (1.5,3)};
\draw [-dot2-=0,black,very thick] ({2-2/sqrt(2)},{2+2/sqrt(2)})-- ({2-2/sqrt(2)},{2+2/sqrt(2)+0.01});

\draw [black,very thick] plot [smooth, tension=1] coordinates {(2,4)  (1.8,3.4)  (1.5,3)};
\draw [-dot2-=0,black,very thick] (2,4)--(2,4.01);

\draw [-dot3-=0.01,-dot3-=0.99,black,very thick, mark position=0.5(a)] plot [smooth, tension=1] coordinates {(2.5,1) (2.1,2.1)   (1.5,3)};
\node at (a) [right] {$\Delta_p$};


\end{tikzpicture}
        \caption{ }
    \end{subfigure}
    \caption{a) Set-up for the calculation of the holographic spinning block. The red points are located at the null boundary at $x_{i=1,2,3,4}$, equipped with  boundary vectors $n^{\partial}_{i}=\partial_u$. The regulated blue points are located at $\gamma_{i}$, and they are equipped with blue vectors $n_{i}=\partial_r\vert_{y_{i}}$. The spinning particles propagate from the blue bulk points to the purple vertex points and also between the two vertex points. At each vertex we have a general time-like vector $n_v$ and $n_{v'}$. The total spinning on-shell action is extremized with respect to the components of these vectors.  b) Alternative way to compute the spinning part of a Poincar\'e block. It is equivalent to the worldline method for computing global conformal blocks in euclidean global AdS$_2$.}\label{fig:4ptspin}
\end{figure}

The set-up for the computation of the spinning Poincar\'e block can be found in Figure \ref{fig:4ptspin}.a). We consider external spins $\Delta$ and internal exchanged spin $\Delta_p$. We start with four regulated  bulk points $y_i$.
These must be connected to two vertices restricted to lie along $N_1\cap N_2$ and $N_3\cap N_4$ respectively, equipped with time-like vectors $n_v$ and $n_{v'}$. The total on-shell action we need to extremize reads 
\beq\label{eq:StotalspinBlock}
\begin{split}
S^{\text{spin}}_{\text{total}} &= -4\Delta\log\epsilon 
+\Delta \left[  \sum^2_{i=1} \log\left(-2 \dot \gamma_i \cdot n_v \right)
+\sum^4_{i=3} \log\left(-2 \dot \gamma_i \cdot n_{v'} \right)
\right]+\Delta_p \cosh^{-1}\left( -n_{v}\cdot n_{v'} \right)\, .
\end{split}
\eeq
Extremizing this object for general boundary angular locations $\phi^{\partial}_i$ is very cumbersome, so we will use BMS transformations to move the operators to the points
\beq
 \phi^{\partial}_{1,2}=\mp \Phi/2\, , \quad  \text{and}\quad \phi^{\partial}_{3,4}=\pi \mp \Phi/2\, .
\eeq
 The cross-ratio $\eta$ is then parametrised by $\Phi$ as
\beq
\eta={{      \sin {{\phi^{\partial}_{12}}\over 2}   \sin {{\phi^{\partial}_{34}}\over 2}          }\over{                  \sin {{\phi^{\partial}_{13}}\over 2}       \sin {{\phi^{\partial}_{24}}\over 2}}}=\sin^2{{\Phi}\over 2}\, .
\eeq
We can now write the bulk vertex vectors in the form \eqref{eq:nvSpin}. Namely,
\beq
\begin{split}
n_v &= {{x^2+u^2+1}\over{2u}}\partial_t +  {{x^2+u^2-1}\over{2u}}\partial_x +{x \over u} \partial_y\, ,\\
n_{v'} &= {{x'^2+u'^2+1}\over{2u'}}\partial_t +  {{x'^2+u'^2-1}\over{2u'}}\partial_x +{x' \over u'} \partial_y\, ,
\end{split}
\eeq
where $u,x$ and $u',x'$ are the variables with respect to which the action \eqref{eq:StotalspinBlock} must be extremal, and as before, they correspond to euclidean AdS$_2$ bulk points in Poincar\'e coordinates. Before continuing it is useful to exploit the symmetry of the problem.  Due to the choice of angular locations of the primary operators, we expect the lengths of the external geodesics to be equal. We can then solve the following system of equations which changes variables from $(u,x,u',x')$ to a single variable $L$
\beq\label{eq:SystemUX}
\begin{split}
-n_i \cdot  n_{v} &= L \sin{{\Phi}\over{2}}\quad \text{for}  \quad i=1,2 \, ,\\
-n_i \cdot  n_{v'} &= L \sin{{\Phi}\over{2}} \quad \text{for}  \quad i=3,4 \, .
\end{split}
\eeq
The solution to \eqref{eq:SystemUX} can be used to write the length between the points parametrised by $n^{\text{bulk}}_{v}$ and $n^{\text{bulk}}_{v'}$ as a function of $L$ exclusively. We thus have a total on-shell action that reads
\beq
\begin{split}
S^{\text{spin}}_{\text{total}} &= \Delta \sum^4_{i=1} \log   {{ L  \sin{{\Phi}\over{2}} }\over{\epsilon}}
+2\Delta_p \log \left(  \cot{{\Phi}\over 2}  {{L\pm \sqrt{L^2-4}}\over 2} \right)\, , \\
&=2\Delta  \log {{\eta}\over {\epsilon}}
+2\Delta_p \log \left( {{1+\sqrt{1-\eta}}\over{\sqrt{\eta}}} \right) + 4\Delta \log L +2\Delta_p \log \left(  L+\sqrt{L^2-4} \right) \, .
\end{split}
\eeq
Extremizing with respect to $L$ gives rise to 
\beq
L={{4\Delta}\over{\sqrt{4\Delta^2-\Delta^2_p}}}\, ,
\eeq
which only contributes to the result by an overall constant that we will ignore. The final result for the $\Delta$-dependent part of the block reads
\beq\label{eq:ResultSpinBlockAction}
e^{-S^{\text{spin}}_{\text{total}}}=C  \eta^{\Delta_p-2\Delta} \left( 1+\sqrt{1-\eta}  \right)^{-2\Delta_p}\, .
\eeq
This matches the large $\Delta$, $\Delta_p$ limit of equations \eqref{eq:Block} and \eqref{eq:4PtFunction}. Note that this result is missing the terms in  \eqref{eq:Block} that are subleading in a large $\Delta$, $\Delta_p$ limit. This is an expected feature of the geodesic network method for computing blocks.

We could have also obtained the same exact result by borrowing results for conformal blocks in CFT$_1$ \cite{Jackiw:2012ur}.   For equal external operator dimensions $\Delta$ and exchanged dimension $\Delta_p$, the CFT$_1$ conformal block decomposition of a four-point function in Poincar\'e time reads
\beq
\langle  \Phi_{\Delta} \Phi_{\Delta} \Phi_{\Delta} \Phi_{\Delta} \rangle_{\text{CFT}_1} = {1\over{(t_{12}t_{34})^{2\Delta}}}\sum_p C_{12p}C_{34p} t^{\Delta_p} {}_2F_1(\Delta_p,\Delta_p;2\Delta_p;t)\, ,
\eeq
with time-like cross ratio 
\beq
t={{t_{12}t_{34}}\over{t_{13}t_{24}}}\, .
\eeq
The map to the boundary of euclidean global AdS$_2$ is written in equation \eqref{eq:Bmap}, and it yields
\beq\label{eq:scalarCFTblock}
\langle   \Phi_{\Delta} \Phi_{\Delta} \Phi_{\Delta} \Phi_{\Delta}  \rangle_{\text{CFT}_1} =  {{    1 }\over{  (\sin{{\phi^{\partial}_{12}}\over 2})^{2\Delta} (\sin{{\phi^{\partial}_{34}}\over 2})^{2\Delta}   }} \sum_p C_{12p}C_{34p}  \eta^{\Delta_p}       {}_2F_1\left( {\Delta_p} , {\Delta_p} ; 2\Delta_p;\eta \right)
\, ,
\eeq
with
\beq
\eta= {{      \sin {{\phi^{\partial}_{12}}\over 2}   \sin {{\phi^{\partial}_{34}}\over 2}          }\over{                  \sin {{\phi^{\partial}_{13}}\over 2}       \sin {{\phi^{\partial}_{24}}\over 2}}} \, .
\eeq
We must now send $\Delta,\Delta_p\rightarrow\infty$. The prefactors multiplying the hypergeometric functions remain unchanged. For the limit of the hypergeometric function  we will use the following identity
\beq\label{eq:2F1}
{}_2F_1(a,b;c;\eta)={{1}\over{B(b,c-b)}}\int^1_0 dt\, t^{b-1}(1-t)^{c-b-1}(1-\eta t)^{-a}\, ,
\eeq
where the beta function also has an integral expression
\beq
B(b,c-b)=\int^1_0 dt\, t^{b-1}(1-t)^{c-b-1}\, .
\eeq
We start with the integral in equation \eqref{eq:2F1}. 
\beq
\begin{split}
{\cal I}&=\int^1_0 dt\, t^{b-1}(1-t)^{c-b-1}(1-\eta t)^{-a}=\int^1_0 dt\, e^{\Delta_p {\cal S}(t)}{1\over{t(1-t)}}\, , \\
{\cal S}(t)&=\log t+\log (1-t)-\log (1-\eta t) \, .
\end{split}
\eeq
We can evaluate this object in a large $\Delta_p$ limit using the method of saddle-point approximation. We need to find the critical points of ${\cal S}(t)$ by solving $\partial_t {\cal S}(t)=0$. There is a unique critical point in the interval $t\in [0,1]$. It reads
\beq
t_{\star}={{1-\sqrt{1-\eta}}\over {\eta}}\, .
\eeq
We thus have
\beq
{\cal I}\rightarrow\sqrt{   {{2\pi}\over {-\Delta_p \partial^2_t{\cal S}(t_{\star})}}  } e^{\Delta_p {\cal S}(t_{\star})} {1\over{t_{\star}(1-t_{\star})}}\, .
\eeq
The integral appearing in the beta function follows a similar logic. 
\beq
\begin{split}
{\cal I}'&=\int^1_0 dt\, t^{b-1}(1-t)^{c-b-1}=\int^1_0 dt\, e^{\Delta_p {\cal S}'(t)}{1\over{t(1-t)}}\, , \\
{\cal S}'(t)&= \log t+\log(1-t)\, .
\end{split}
\eeq
Solving for $\partial_t {\cal S}'(t)=0$ results in a unique critical point at
\beq
t'_{\star}={1\over 2}\, .
\eeq
We conclude
\beq
{\cal I}'\rightarrow\sqrt{   {{2\pi}\over {-\Delta_p \partial^2_t{\cal S}'(t'_{\star})}}  } e^{\Delta_p {\cal S}'(t'_{\star})} {1\over{t'_{\star}(1-t'_{\star})}}\, ,
\eeq
which implies, after some simple algebra
\beq\label{eq:2F1Final}
{}_2F_1\left({\Delta_p},{\Delta_p};2\Delta_p;\eta\right)\rightarrow   2^{2\Delta_p-1}(1-\eta)^{-{1\over 4}} \left( 1+\sqrt{1-\eta}  \right)^{1-2\Delta_p}  \, .
\eeq
Replacing this result in \eqref{eq:scalarCFTblock} we obtain
\beq\label{eq:resultblockspin}
\langle  \Phi_{\Delta}\Phi_{\Delta}\Phi_{\Delta}\Phi_{\Delta} \rangle_{\text{CFT}_1} = {{    1      }\over{  (\sin{{\phi^{\partial}_{12}}\over 2})^{2\Delta} (\sin{{\phi^{\partial}_{34}}\over 2})^{2\Delta}   }}   \sum_p 
 C_{12p}C_{34p} 2^{2\Delta_p} \eta^{\Delta_p} \left( 1+\sqrt{1-\eta}  \right)^{-2\Delta_p} 
\, ,
\eeq
where we have ignored the subleading factors. The answer matches the result \eqref{eq:ResultSpinBlockAction} obtained from extremizing the spinning on-shell action.  It also agrees with the large $\Delta, \, \Delta_p$ limit of equations \eqref{eq:4PtFunction} and \eqref{eq:Block}, which were obtained in the BMSFT in \cite{Bagchi:2017cpu}. 

Note that the factors obtained in this section that are subleading in the large $\xi$, $\Delta$ limit do not match the correct result as presented in equation \eqref{eq:Block}. This is an expected feature of the worldline method, which applies exclussively in the probe limit. In order to obtain the full answer we need to build a full extrapolate dictionary and use it to make a proposal for the holographic computation of the block in the light regime.

\section{A proposal for the extrapolate dictionary}\label{sec:extrapolate}
Given the results presented in this note, it is natural to ask whether the construction we have proposed above can be understood as the probe limit of a full extrapolate dictionary. In this section, we propose such a dictionary and use it to match the above results in a WKB approximation. By computing corrections to this approximation, we will be able to match the finite dimension form of BMS$_3$ correlation functions and blocks. 

The  prescription presented throughout this work consists of attaching a geodesic network to a set of null lines falling from the boundary from the locations of the boundary insertions. Given a point $x$ at the boundary, we thus associate a unique null line $\gamma_x$. The geodesics that constitute the network correspond to the saddle points of flat space propagators. The proposed extrapolate dictionary is then to attach a position space Feynman diagram to the null lines and integrate the position of the legs over an affine parameter along $\gamma_x$.
\begin{align}\label{eq:Extrapolate}
\langle \cO_1(x_1) \cO_2(x_2) \ldots \rangle 
\sim \int_{\gamma_{x_1}} d\lambda_1 \int_{\gamma_{x_2}} d\lambda_2 \ldots 
\langle \Psi_1(\lambda_1) \Psi_2(\lambda_2) \ldots \rangle\, .
\end{align}
Here, $\lambda_i$ is  the affine parameter along the null geodesic $\gamma_{x_i}$ and $\Psi_i$ is the bulk field dual to the operator  $\cO_i$. The symbol $\sim$ is used because both sides can be rescaled by arbitrary constant factors as both the normalisation of the operators and the measure of an affine parametrisation can be rescaled. We might have used a different measure when integrating the bulk operators along the null geodesics, but we will find that this choice reproduces the correct BMS$_3$ correlation functions.

The objective of this section is to test formula \eqref{eq:Extrapolate} for the case of propagating massive scalar fields in Minkowski space-time. We expect to compute the exact form of correlators involving boundary operators with mass $m=\xi$ and spin $\Delta=1$, as explained in section \ref{sec:GravityIntro}. 

\subsection{Two-point function}
The easiest correlation on which formula \eqref{eq:Extrapolate} can be tested is the two point function. The object to compute reads
\begin{align}
\langle \cO (x_1) \cO(x_2)\rangle 
\sim \int_{\gamma_{x_1}} d\lambda_1 \int_{\gamma_{x_2}} d\lambda_2 \, 
\langle \Psi (\lambda_1) \Psi (\lambda_2)  \rangle \, ,
\end{align}
where the integrand is simply the bulk to bulk Feynman propagator for the field dual to ${\cal O}$. For the case of a scalar field, it is useful to write this object in momentum space form
\beq
\langle \Psi (\lambda_1) \Psi (\lambda_2)  \rangle =G_F (\lambda_1,\lambda_2)= \int d^3p  {{e^{i p_{\mu} (x^{\mu}_{\gamma_1}-x^{\mu}_{\gamma_2})}}\over{p_{\mu}p^{\mu}+\xi^2 -i \epsilon}}\, ,
\eeq
where $x^{\mu}_{\gamma_i}$ are the locations of the null lines
\beq
x^{0}_{\gamma_i} = u_i + \lambda_i\, , \quad x^{1}_{\gamma_i}=\lambda_i \cos \phi^{\partial}_i\, , \quad \text{and} \quad x^{2}_{\gamma_i}=\lambda_i \sin \phi^{\partial}_i\, ,
\eeq
and the components of the three-momentum are dubbed as $p^{\mu}=\{  w,p^x,p^y \}$.

The calculation of the correlator can be started by performing the integrals over the affine parameters $\lambda_i$. The relevant integrals read
\beq\label{eq:delta}
\int d\lambda_i e^{i \lambda_i (-w+p^x \cos\phi^{\partial}_i+p^y \sin\phi^{\partial}_i)} = 2\pi \delta\left(-w+p^x \cos\phi^{\partial}_i+p^y \sin\phi^{\partial}_i\right)\, .
\eeq
We can thus inmediately integrate the momenta $p^x$ and $p^y$, which in virtue of the delta functions are forced to correspond to the following expressions
\beq
p^x=w{{\sin\phi^{\partial}_{1}-\sin \phi^{\partial}_{2} }\over{\sin\phi^{\partial}_{12}}}\, , \quad p^y=-w{{\cos\phi^{\partial}_{1}-\cos \phi^{\partial}_{2} }\over{\sin\phi^{\partial}_{12}}}\, .
\eeq
The remaining integral over $w$ is
\beq
\langle \cO_1(x_1) \cO_2(x_2) \rangle \sim {1\over {\sin\phi^{\partial}_{12} }}\int dw {e^{-i w u_{12} }\over{w^2 \tan^2{{\phi^{\partial}_{12}}\over 2}+\xi^2 -i \epsilon}}\, .
\eeq
A simple Fourier transform yields the right result
\beq
\langle \cO_1(x_1) \cO_2(x_2) \rangle \sim {{1}\over{\sin^2{{\phi^{\partial}_{12}}\over 2} }} e^{\xi {{u_{12}}\over{\tan{{\phi^{\partial}_{12}}\over 2}}} }\, .
\eeq
The calculation done in this section arives to the same result obtained in previous sections using the probe limit. However, the method used here does not rely on a large mass limit. This will be more obvious when studying Poincar\'e blocks, as the probe limit does not capture the exact result for the block. Before we study these issues, we turn our attention to the next more complicated correlator; the three-point function.

\subsection{Three-point function}
The calculation of the three-point correlator is morally simmilar to the computation of the two-point function, albeit the addition of a bulk vertex where the propagators interact through a coupling of the form $\Psi^3$. The computation we propose is to attach a tree level Feynman diagram to the null geodesics following from the boundary. We thus have
\begin{align}\label{eq:deltalambda}
\langle \cO_1(x_1) \cO_2(x_2)\cO_3(x_3)\rangle 
\sim \int_{\gamma_{x_1}} d\lambda_1 \int_{\gamma_{x_2}} d\lambda_2 \int_{\gamma_{x_3}} d\lambda_3\, 
\langle \Psi_1(\lambda_1) \Psi_2(\lambda_2)\Psi_3(\lambda_3)  \rangle \, ,
\end{align}
where the bulk correlator reads
\begin{align}
\langle \Psi_1(\lambda_1) \Psi_2(\lambda_2)\Psi_3(\lambda_3)  \rangle \sim \int d^3 y_v  \prod^3_{i=1} \int d^3p_i {{e^{i p_{i,\mu} (x^{\mu}_{\gamma_i}-y^{\mu}_{v})}}\over{p_{i,\mu}p_i^{\mu}+\xi_i^2 -i \epsilon}} \, .
\end{align}
The Feynman propagators connect points in the null geodesics $\gamma_i$ with a common bulk vertex at $y_v$, which must be integrated over all of Minkowski space-time in order to obtain a gauge invariant answer. As we did in the previous subsection, we start by integrating over the null geodesics. Again the relevant integrals are the ones in equation \eqref{eq:deltalambda}. The delta functions can be used to integrate the variables $p^0_i=w_i$, which are constrained to the values 
\beq
w_i=p_i^x \cos\phi^{\partial}_i+p_i^y \sin\phi^{\partial}_i\, .
\eeq
After performing the integrals over $w_i$, it is useful to change variables as follows
\beq\label{eq:ChangeP}
\tilde{p}_i=p_i^x \sin \phi^{\partial}_i-p_i^y \cos \phi^{\partial}_i \, .
\eeq
The correlator then reads
\beq
\begin{split}
\langle \Psi_1(\lambda_1) \Psi_2(\lambda_2)\Psi_3(\lambda_3)  \rangle\sim& \int d^3 y_v \prod_i \int d\tilde{p}_i  {{e^{-i\tilde{p}_i  {{x_v+(u_i-t_v)\cos\phi^{\partial}_i}\over{\sin\phi^{\partial}_i}} }}\over{\tilde{p}_i^2+\xi_i^2-i\epsilon}}\int dp_i^y {{e^{-i p_i^y {{u_i-t_v+r_v \cos(\phi_v-\phi^{\partial}_i) }\over{\sin \phi^{\partial}_i}}}}\over{\sin \phi^{\partial}_i}} \, .
\end{split}
\eeq
The integrals over $p_i^y$ are delta functions that completely fix the vertex point $y_v$. The expression for the resulting coordinate location of the vertex is precisely the expression \eqref{eq:xv}, which was found in previous sections studying geodesic networks as the probe limit of \eqref{eq:Extrapolate}. The remaining integrals over $\tilde{p}_i$ are simple Fourier transforms. The final result reads
\beq
\langle \Psi_1(\lambda_1) \Psi_2(\lambda_2)\Psi_3(\lambda_3)  \rangle\sim\prod_k{{e^{
\xi_k{{\sum_{i<j}(-1)^{1+i+j}(u^{\partial}_i-u^{\partial}_j)\cos(\phi^{\partial}_k-\phi^{\partial}_i)}
\over
{\sum_{i<j}(-1)^{1+i+j}\sin(\phi^{\partial}_i-\phi^{\partial}_j)}}
} }} \left( \sin {{\phi^{\partial}_i-\phi^{\partial}_j}\over 2} \right)^{-1}\, ,
\eeq
which matches \eqref{eq:lowpt} upon replacing $\Delta_i = 1$. Much like the case of the two point function, the result here matches the result obtained from studying geodesic networks.

\subsection{Poincar\'e Blocks - geodesic Feynman diagrams}\label{sec:GFD}
 We now explore  holographic Poincar\'e blocks following the extrapolate dictionary proposed in this section. The new feature of this calculation is the fact that the four-point correlator $\langle \Psi_1(\lambda_1) \Psi_2(\lambda_2)\Psi_3(\lambda_3) \Psi_4(\lambda_4)  \rangle$ is not fixed by Lorentz invariance.  We could think of attaching any tree level Feynman diagram to the null lines $\gamma_i$ to compute gauge invariant objects. The work in \cite{Hijano:2015rla,Hijano:2015zsa} showed that, in the context of the AdS/CFT correspondence, holographic conformal blocks are computed with geodesic Witten diagrams (GWD). These objects are defined as exchange Witten diagrams where the internal vertices are not integrated over the whole bulk manifold, but only over the geodesics connecting pairs of boundary operators. Inspired by the AdS picture, we now construct a geodesic Feynman diagram (GFD) by integrating the vertices of an exchange Feynman diagram over the geodesics $\gamma_{12}=N_1\cap N_2$ and $\gamma_{34}=N_3\cap N_4$. As we have seen above, these geodesics are the paths of extremal length connecting the points at the null boundary. We parametrise these geodesics in terms of their arc-length as follows
 \beq
 \begin{split}
   t_{\gamma_{12}}&=s \cot {{\phi^{\partial}_{21}}\over 2}\, , \\
 x_{\gamma_{12}}&=-s{{\cos\phi^{\partial}_2+\cos\phi^{\partial}_1}\over{\sin\phi^{\partial}_{21}}}+{{u^{\partial}_2\sin\phi^{\partial}_1-u^{\partial}_1\sin\phi^{\partial}_2}\over{\sin\phi^{\partial}_{21}}} \, , \\
 y_{\gamma_{12}}&=-s {{\sin\phi^{\partial}_2+\sin\phi^{\partial}_1}\over{\sin\phi^{\partial}_{21}}} +{{u^{\partial}_1\cos\phi^{\partial}_2-u^{\partial}_2\cos\phi^{\partial}_1}\over{\sin\phi^{\partial}_{21}}} \, ,
\end{split}
 \eeq
and simmilarly for $\gamma_{34}$. The holographic block as a GFD attached to the null geodesics thus reads
\beq\label{eq:GFD}
\begin{split}
\langle \cO_1(x_1) \cO_2(x_2)\cO_3(x_3)\cO_4(x_4)\rangle 
\sim& \int_{\gamma_{12}} ds \int_{\gamma_{34}} ds' \left(\prod_i \int_{\gamma_{x_i}} d\lambda_i\right)  \, \\
&G^{\xi}_F (\lambda_1,\sigma)G^{\xi}_F (\lambda_2,\sigma)G^{\xi_p}_F(\sigma,\sigma')G^{\xi}_F (\lambda_3,\sigma')G^{\xi}_F (\lambda_4,\sigma')\, ,
\end{split}
\eeq
where $G_F^{m}$ stand for Feynman propagators of spinless fields of mass $m$. For simplicity, we will choose the operators at the null boundary to be placed at 
\beq
\begin{split}
\phi^{\partial}_1&=-{{\Phi}\over 2}\, , \quad \phi^{\partial}_2={{\Phi}\over 2}\, , \quad\phi^{\partial}_3=\pi-{{\Phi}\over 2}\, , \quad\phi^{\partial}_4=\pi+{{\Phi}\over 2}\, , \\
u^{\partial}_1&=-{T\over 2}\, , \quad u^{\partial}_2={T\over 2}\, , \quad u^{\partial}_3=-{T\over 2}\, , \quad u^{\partial}_4={T\over 2}\, .
\end{split}
\eeq
This leads to the cross ratios
\beq\label{eq:CR}
x=\sin^2 {{\Phi}\over 2}\, , \quad t=T \sin {{\Phi}\over 2} \cos {{\Phi}\over 2}\, .
\eeq

The calculation of \eqref{eq:GFD} can be done by writing the Feynman propagators in momentum space form. For the external propagators, we use the notation $p^{\mu}_i=\left( w_i,p_i^x,p_i^y\right)$, while for the internal propagator we use $p_p^{\mu}=\left(w_p,p_p^x,p_p^y \right)$. The integrals over $\lambda_i$ can then be done inmediately, resulting on delta functions of the form \eqref{eq:delta}. The resulting expression can then be simplified by changing variables as in equation \eqref{eq:ChangeP}. The integral over the arc-length $s$ then reads
\begin{align}
\int_{\gamma_{12}} ds e^{-i s \left(  \tilde{p}_1-\tilde{p}_2-w_p\cot{{\Phi}\over 2} +p_p^x \csc{{\Phi}\over 2} \right) }=2\pi  \delta\left(  \tilde{p}_1-\tilde{p}_2-w_p\cot{{\Phi}\over 2} +p_p^x \csc{{\Phi}\over 2}    \right) \, ,
\end{align}
and similarly for the integral over $s'$. These delta functions fix $w_p$ and $p_p^x$ as a function of $\tilde{p}_i$. We are left with integrals over $p_p^y$, $p_i^y$, and $\tilde{p}_i$. The resulting integrand however does not depend on $p_i^y$, so the corresponding integrals contribute to an overall infinity. The remaining calculation reads
\beq\label{eq:remaining}
\begin{split}
\langle \cO_1(x_1) \cO_2(x_2)\cO_3(x_3)\cO_4(x_4)\rangle 
\sim&  {{\cos {{\Phi}\over 2}}\over{\sin^2\Phi}} \left(\prod_i \int d\tilde{p}_i {{e^{-i\tilde{p}_i{T\over 2}\cot{{\Phi}\over 2}}}\over{\tilde{p}_i^2+\xi^2-i\epsilon}}\right)
\int dp^y_p{{e^{-ip^y_p {T\over{\sin{{\Phi}\over 2}}}}}\over{
A(p_p^y, \tilde{p}_i)
}}\, , 
\end{split}
\eeq
with 
\beq
A(p_p^y, \tilde{p}_i)= (p_p^y)^2+\xi_p^2 +{{\sin^2{{\Phi}\over 2}}\over 4}(\tilde{p}_1-\tilde{p}_2+\tilde{p}_3-\tilde{p}_4)^2-{{\tan^2{{\Phi}\over 2}}\over 4} (\tilde{p}_1-\tilde{p}_2-\tilde{p}_3+\tilde{p}_4)^2\, .
\eeq
We now proceed by using a Schwinger identity for the denominator in the $p_p^y$ integral of \eqref{eq:remaining}. 
\beq
{1\over {A(p_p^y, \tilde{p}_i)}}=\int_0^{\infty} d\nu e^{-\nu A(p_p^y, \tilde{p}_i)}\, .
\eeq
The resulting expression is easy to integrate with respect to the variables $\tilde{p}_i$ using the residue theorem. Namely, due to the choice of Feynman contour, only the residues $\tilde{p}_i=+i\xi$ contribute to the final result. Computing the residues yields
\beq\label{eq:remainingv2}
\begin{split}
\langle \cO_1(x_1) \cO_2(x_2)\cO_3(x_3)\cO_4(x_4)\rangle 
\sim {{\cos {{\Phi}\over 2}}\over{\sin^2\Phi}} e^{2\xi T \cot{{\Phi}\over 2}} \int_0^{\infty} d\nu\int dp_p^y e^{-\nu ((p_p^y)^2+\xi_p^2)} e^{-i p_p^y {T\over{\sin{{\Phi}\over 2}}}}\, .
\end{split}
\eeq
The integral over the Schwinger parameter now yields
\beq\label{eq:remainingv3}
\begin{split}
\langle \cO_1(x_1) \cO_2(x_2)\cO_3(x_3)\cO_4(x_4)\rangle 
\sim {{\cos {{\Phi}\over 2}}\over{\sin^2\Phi}} e^{2\xi T \cot{{\Phi}\over 2}}\int dp_p^y  {{   e^{-i p_p^y {T\over{\sin{{\Phi}\over 2}}}}}\over{(p_p^y)^2+\xi_p^2}}\, .
\end{split}
\eeq
The last integral over $p_p^y$ is a simple Fourier transform, which leads to the final result 
\beq\label{eq:remainingv4}
\begin{split}
\langle \cO_1(x_1) \cO_2(x_2)\cO_3(x_3)\cO_4(x_4)\rangle 
\sim {1\over{\cos{{\Phi}\over 2}\sin^2{{\Phi}\over 2}}} e^{2\xi T \cot{{\Phi}\over 2}-\xi_p {T\over{\sin{{\Phi}\over 2}}}}\, .
\end{split}
\eeq
In terms of the cross ratios of equation \eqref{eq:CR} we obtain
\beq\label{eq:final}
\begin{split}
\langle \cO_1(x_1) \cO_2(x_2)\cO_3(x_3)\cO_4(x_4)\rangle 
\sim  x^{-1}(1-x)^{-{1\over 2}} e^{2\xi {t\over x}-\xi_p {t\over {x\sqrt{1-x}}}}\, .
\end{split}
\eeq
This results matches exactly the result \eqref{eq:Block} upon the replacement $\Delta_p=\Delta_i=1$. Note that the prefactors which are subleading in the large $\xi$, $\xi_p$, $\Delta$, $\Delta_p$ limit were not obtained in the probe limit of section \ref{sec:SPIN}. The computation in this section does generate the correct result, as it should be valid for any value of the mass and the spin of the bulk fields, as long as these do not backreact the background geometry. We conclude that the extrapolate dictionary proposed in \eqref{eq:Extrapolate}, together with the introduction of geodesic Feynman diagrams, provide a holographic picture for Poincar\'e blocks in three dimensional Minkowski space.

\section{Discussion}
In this note we have proposed an extrapolate dictionary in the context of Einstein gravity in three dimensional Minkowski space-time, dual to a BMS$_3$ field theory at the null boundary. We have tested this dictionary through the holographic computation of BMS$_3$ entanglement entropy, low-point correlators, and global blocks. We have further introduced the concept of geodesic Feynman diagrams, in imitation of geodesic Witten diagrams in the AdS/CFT context. In the probe limit, this construction leads to the extremization of the length of a network of straight lines attached to the asymptotic boundary through null geodesics. There are a number of inmediate directions of future research. Let us discuss some of them.

The extrapolate dictionary \eqref{eq:Extrapolate} has been tested for scalar fields propagating through flat space. It would also be interesting to test its validity in the case of propagating spinning particles, which should lead to correlation functions involving boundary operators with quantum numbers $\Delta\neq1$. For the case of spinning propagators, the prescription for the bulk-to-boundary propagator should correspond to integrating the pullback of the Feynman propagator to the null line $\gamma_i$ falling from the boundary point where the corresponding operator is inserted. 

The holographic construction of conformal blocks in AdS was first studied in \cite{Hijano:2015zsa}, where the relation between Witten diagrams and conformal blocks was also explored. In the case of CFT scalar operators, it was shown that Witten diagrams and global blocks are distinct basis of conformally invariant functions. The authors of \cite{Hijano:2015zsa} were also able to expand a single Witten diagram into the global block basis. Denoting by $W_p$ an exchange Witten diagram exchanging the conformal dimension $h_p$ with pairs of operators of conformal dimensions $h$, the result reads
\beq
W_p={\cal W}_p+\sum_n a_n {\cal W}_n +\sum_m a_m {\cal W}_m \, ,
\eeq
where ${\cal W}_p$ is the global block exchanging the representation labeled by $h_p$, while ${\cal W}_{n,m}$ correspond to blocks exchanging conformal families associated to double trace operators built out of the external operator pairs.  In the context of flat space, one can also wonder what the relation is between attaching a standard position space Feynman diagram to the null lines $\gamma_i$ ($F_p$), an attaching a geodesic Feynman diagram as done in section \ref{sec:GFD} (${\cal F}_p$). A simple calculation shows that for the case of bulk scalar fields, the geodesic Feynman diagram leads to the same result as the standard Feynman diagram
\beq
F_p={\cal F}_p \, .
\eeq
This implies that the Feynman diagram basis is equivalent to the global BMS block basis. It is not clear to us what this means in terms of the nature of BMS field theories, but it clearly points to a fundamental difference between holography in flat space and holography in anti-de Sitter space.

It would also be interesting to explore flat space holography in higher dimensions. In \cite{Bagchi:2016bcd}, the two- and three-point functions invariant under BMS$_4$ were computed. It was found that the answer matched the expression for correlators in a two dimensional conformal field theory, which hints at the statement that gravity in Minkowski$_4$ space-time might be dual to a relativistic CFT$_2$.

\acknowledgments
The authors gratefully acknowledge Wei Song, Arjun Bagchi, Tarek Anous, Felix Haehl, Eric Mintun, Mark Van Raamsdonk, Robert Penna, Shu-Heng Shao and Prahar Mitra for very useful discussions. 

We wish to thank  Centro de Ciencias de Benasque Pedro Pascual for hospitality during this project. The work of  EH is supported in part by the Natural Sciences and Engineering Research Council of Canada,  and the “It From Qubit” collaboration grant from the Simons Foundation. 
CR wishes to acknowledge support from the Simons Foundation (\#385592, Vijay Balasubramanian) through the It From Qubit Simons Collaboration, from the Belgian Federal Science Policy Office through the Interuniversity Attraction Pole P7/37, by FWO-Vlaanderen through projects G020714N and G044016N, and from Vrije Universiteit Brussel through the Strategic Research Program ``High-Energy Physics''.

\bibliographystyle{JHEP}
\bibliography{refs}
\end{document}